\documentclass[fleqn,usenatbib]{mnras}

\usepackage{newtxtext,newtxmath}

\usepackage[T1]{fontenc}
\usepackage{ae,aecompl}

\usepackage{graphicx}	
\usepackage{amsmath}	
\usepackage{amssymb}	
\usepackage{multirow}
\usepackage{booktabs}
\usepackage{color}

\newcommand{\Msun}{\ifmmode {M_{\odot}}\else${M_{\odot}}$\fi}

\newcommand{\cory}[1]{{\color{black}{#1}}}
\newcommand{\frcory}[1]{{\color{black}{#1}}}

\newcommand{\mujy}{\mathrm{\mu Jy}}

\title[Radio sources of NGC 6397]{The MAVERIC survey: A hidden pulsar and a black hole candidate in ATCA radio imaging of the globular cluster NGC 6397}

\author[Yue Zhao et al.]{
Yue Zhao,$^{1}$\thanks{E-mail: zhao13@ualberta.ca}
Craig O. Heinke,$^{1}$
Vlad Tudor,$^{2}$
Arash Bahramian,$^{2}$
\newauthor
James C. A. Miller-Jones$^{2}$,
Gregory R. Sivakoff,$^{1}$
Jay Strader,$^{3}$
Laura Chomiuk,$^{3}$
\newauthor
Laura Shishkovsky,$^{3}$
Thomas J. Maccarone,$^{4}$
Manuel Pichardo Marcano,$^{4}$ 
\newauthor
and Joseph D. Gelfand$^{5,6}$
\\
$^{1}$ Department of Physics, University of Alberta, CCIS 4-183, Edmonton, AB, T6G 2E1, Canada\\
$^{2}$ International Centre for Radio Astronomy Research-Curtin University, GPO Box U1987, Perth, WA 6845, Australia \\
$^{3}$ Department of Physics and Astronomy, Michigan State University, East Lansing, MI 48824, USA \\
$^{4}$ Texas Tech University, Dept. of Physics, Box 41051, Lubbock, TX, 79409-1051, USA \\
$^{5}$ New York University Abu Dhabi, Abu Dhabi, United Arab Emirates \\
$^{6}$ NYU Center for Cosmology and Particle Physics, New York, NY 10003, USA
}

\date{Accepted XXX. Received YYY; in original form ZZZ}

\pubyear{2019}

\begin{document}
\label{firstpage}
\pagerange{\pageref{firstpage}--\pageref{lastpage}}
\maketitle

\begin{abstract}
Using a $16.2~\mathrm{hr}$ radio observation by the Australia Telescope Compact Array (ATCA) and archival {\it Chandra} data, we found 
$>5\sigma$ 
radio counterparts 
to $4$ known and $3$ new X-ray sources within the half-light radius ($r_\mathrm{h}$) of the Galactic globular cluster NGC 6397. The previously suggested millisecond pulsar (MSP) candidate, U18, 
is a steep-spectrum
  ($S_\nu\propto \nu^\alpha$; $\alpha=-2.0^{+0.4}_{-0.5}$) radio source with a $5.5$ GHz flux density of $54.7\pm 4.3~\mu\mathrm{Jy}$. 
  We argue that U18 is most likely a ``hidden" MSP that is continuously hidden by plasma shocked at the collision betwen the winds from the pulsar and companion star. 
  The nondetection of radio pulsations so far is probably the result of enhanced
  scattering in this shocked wind.
  On the other hand, we observed the 5.5 GHz flux of the known MSP PSR J1740-5340 (U12) to decrease by a factor of $>$2.8 during epochs of 1.4 GHz eclipse, indicating that the radio flux is absorbed in its shocked wind. 
  \frcory{If U18 is indeed a pulsar whose pulsations are scattered, we note the contrast with U12's flux decrease in eclipse, which argues} for two different eclipse mechanisms at the same radio frequency.
  \frcory{In addition to U12 and U18}, we also found radio associations for 5 other {\it Chandra} X-ray sources, four of which are likely background galaxies. The last, U97, which shows strong $\mathrm{H}\alpha$ variability, is mysterious; it may be either a quiescent black hole low-mass X-ray binary, or something more unusual.
\end{abstract}




\section{Introduction}
X-ray observations have revealed that Galactic globular clusters (GCs) host an overabundance of X-ray sources, which are thought to originate from compact binaries formed through close encounters in the dense cluster cores \citep{Fabian75, Hills76, bailyn1990, camilo2005, ivanova2006}. One class of dynamically formed close binaries consists of low-mass X-ray binaries (LMXBs), where neutron stars (NSs) accrete matter from low-mass companion stars \citep{lewin1983, grindlay1984}. These LMXBs are the progenitors of millisecond pulsars (MSPs), radio pulsars with very stable (\frcory{low $\dot P$}) millisecond-scale spin periods ($P\sim1-10~\mathrm{ms}$) and low \frcory{spindown-inferred dipole} magnetic fields ($B\sim10^{8-9}~\mathrm{G}$) \citep{Manchester17}, which are also abundant in globular clusters \citep[e.g.][]{camilo2005}. Cataclysmic variables (CVs), white dwarfs (WDs) accreting from low-mass stars, can also be produced dynamically in dense globular clusters \citep{pooley2003,Heinke03d,Pooley06,ivanova2006}, but in many clusters most CVs are primordial in origin \citep{Verbunt88,Davies97,Kong06,Bassa08,Haggard09,Cheng18,Belloni19}. Chromospherically active binaries (ABs), which are close binaries involving rapidly rotating late (K-M) type (BY Draconis) or evolved stars (RS CVn), make up the majority of faint X-ray sources below $10^{31}~\mathrm{erg~s^{-1}}$, and are mostly formed primordially \citep{grindlay2001,Bassa04}.
\\

\subsection{Radio sources in globular clusters}
\label{sec:gc_radio_src}
Thanks to the increased sensitivity of new {\it Karl G. Jansky Very Large Array} (VLA) and ATCA receivers, faint radio sources have been detected in a number of GCs \citep[e.g.][]{Strader12,Chomiuk13,MillerJones15}. There are several possibilities for these sources.

Neutron star LMXBs in the low/hard states produce flat \frcory{($-0.5<\alpha<0$; $S_\nu \propto \nu^\alpha$)} to inverted \frcory{($\alpha \geq 0$)} radio spectra, thought to be from jets, correlated with their X-ray luminosity \citep{Migliari06,Tudor17,Gusinskaia19}. Black hole LMXBs produce jets with similar radio spectra, but \frcory{the radio luminosity is significantly higher} for a given X-ray luminosity \citep{Gallo14, Gallo18, Plotkin19}. Millisecond radio pulsars generally show steep spectral indices, with a mean of $\alpha= -1.4$ and unit standard deviation \citep{Bates13}.  

Non-magnetic\footnote{CVs are divided into magnetic systems, where the WD magnetic field is strong enough (\frcory{$B$ $\sim 10^{6-8}~\mathrm{G}$}) to channel the accretion onto magnetic poles, and non-magnetic systems where an accretion disc reaches the surface.} CVs show flat to inverted radio spectra during accretion episodes, believed to be synchrotron emission from a jet \citep{Benz83,Kording08,MillerJones13,Coppejans15,Coppejans16, Russell16}. 
Numerous magnetic CVs have been detected in the radio, though a number of magnetic CVs show circular polarization indicative of electron-cyclotron maser emission \citep{Abada-Simon93,Chanmugam87,Barrett17}. This maser emission has been suggested to be produced near the companion star \citep{MasonGray07}, or to be produced near the WD \citep{Kurbatov19}. Radio luminosities for all CVs tend to be rather low \frcory{(10s to 100s of $\mu$Jy at 300 pc; e.g. \citealt{Coppejans16})}, except during brief flares within dwarf nova outbursts \citep{Mooley17}.
The most radio-luminous WDs known are the extreme magnetic CV AE Aquarii, which is thought to eject much of the infalling matter through a propeller \citep{Wynn97,Meintjes05}, and the WD pulsar AR Sco, where the WD does not accrete, but produces nonthermal emission from the radio through X-ray \citep{Marsh16}.

ABs are also radio sources, though generally they are not \frcory{luminous} enough for detection at kpc distances except during short (hours) flares \citep{Drake89,Osten00}. Tidal interaction in close orbits leads to synchronised rotation with short orbital period and thus strong coronal magnetic activity \citep{Chugainov66, bailyn1990, dempsey1993}.
The emission mechanism is generally attributed to gyrosynchrotron radiation of mildly relativistic electrons interacting with photospheric magnetic fields \citep{Hjellming80, Feldman83, Guedel92, Kundu85}. They are generally observed to be non-thermal, highly variable and circularly polarised with flat or negative spectral indices ($\alpha\lesssim 0$; \citealt{Garcia03}).
An unusual system, a sub-subgiant\footnote{A star fainter than subgiants but redder than the main sequence; see \citealt{Leiner17}} in a binary with either a very low-mass star, or else with a more massive compact object (possibly a BH) in an extremely face-on binary, is a radio and X-ray source in the globular cluster M10 \citep{Shishkovsky18}. 

\subsection{Eclipsing millisecond pulsars}
\label{sec:sec_eclipsing_msps}
MSPs are often in close orbits with very low-mass (as low as $\sim 0.02~M_\odot$) stars. 
\frcory{If the companion is not fully degenerate, it may produce an outflowing wind, which interacts with the pulsar wind to create an intrabinary shock.
The pulsar's radiation and/or accelerated particles may heat the companion's surface, enhancing the companion's wind. This enhanced companion wind, and/or the intrabinary shock, may eclipse the radio pulsations.}
These systems fall into two major categories, ``black widow" systems with brown dwarf companions of $\lesssim 0.02$ \Msun  \citep{Fruchter88,Stovall14}, and ``redback" systems with main sequence companions of typically $\sim 0.1$-$0.4$ \Msun \citep{Lyne90,Ferraro01,Roberts13}\frcory{, but possibly higher companion masses up to $0.7$-$0.9~M_\odot$\citep{Strader19}.} The radio eclipses are often seen to encompass 10-25\% of the full orbit, but the eclipse lengths \frcory{can} vary between orbits, and at low frequencies some systems appear to be permanently eclipsed \citep[e.g.][]{Camilo00,Freire05}. 
There are also MSPs such as 47 Tuc V which show  irregular eclipses at all orbital phases and are sometimes not detected for many orbits, suggesting they are continuously eclipsed \citep{Camilo00,Ridolfi16}. It is speculated that  many MSPs may be continuously hidden behind even stronger winds \citep{Tavani91}.

The mechanism by which the wind from the companion star causes eclipses in the radio pulsations is not clear. The radio pulses typically show increased dispersion near the eclipse, and/or become substantially fainter \citep{Stappers01,Archibald09}.  \citet{Thompson94} discuss a range of possible mechanisms, including free-free absorption, pulse smearing, scattering due to Langmuir turbulence, stimulated Raman scattering, and cyclotron absorption. \citet{Thompson94} and later works \citep{Stappers01,Polzin19} typically favor cyclotron absorption and/or scattering mechanisms, while some works \citep{Broderick16,Polzin18} strongly favor cyclotron absorption to explain eclipses, especially at low (e.g. 300 MHz) frequencies. \citet{FruchterGoss92} imaged PSR 1957+20 through eclipses at both 20 and 90 cm with the VLA, discovering that the unpulsed flux disappeared during eclipse at 90 cm but was still present at 20 cm, suggesting cyclotron absorption at low frequencies and scattering at higher frequencies \citep{Thompson94}. The LOFAR array has been used to image pulsars during eclipses, where the disappearance of the flux at e.g. 149 MHz was seen, in agreement with cyclotron absorption scenarios \citep{Roy15,Broderick16,Polzin18}.

Observations of lengthy X-ray eclipses in redbacks were initially understood as a direct eclipse by the secondary star \frcory{of X-ray emission from} an intrabinary shock, located close to the secondary \citep{Bogdanov05}. However, \frcory{higher S/N} orbital X-ray lightcurves of eclipsing pulsars have revealed modulation of the X-rays throughout the orbit, and sharp peaks, often on either side of the inferior conjunction of the NS \citep[e.g.][]{Bogdanov11,Romani11,Huang12,Bogdanov14,Hui14,deMartino15,Hui15}. This has inspired interpretation of the X-rays as due to particle acceleration at the interface between the companion and pulsar winds, beamed in the direction of the particle flow \citep[e.g.][]{Harding90,Arons93,Romani16,Wadiasingh17}. The very hard X-ray photon index rules out shock acceleration, suggesting magnetic reconnection in a striped pulsar wind \citep[e.g.][]{alNoori18}.  It is unclear how the intrabinary shock manages to wrap around the pulsar; suggested scenarios are that the companion wind is highly magnetized (as the companion is likely magnetically active, \citealt{vanStaden16}) and thus that the companion wind balances the pulsar's via magnetic pressure, or that the companion wind is dense enough \frcory{for its gas pressure to balance the pulsar wind's magnetic pressure} \citep{Wadiasingh18}. The latter case is inherently unstable to gravity if the intrabinary shock bends around the pulsar, which may explain rapid transitions between accretion and pulsar states \citep{Papitto13}. We note that \citet{Li18} \frcory{place very constraining upper limits on} the $B$ field at the interface betwen the pulsar and companion winds in PSR B1957+20, casting doubt on the cyclotron absorption eclipse scenario and the magnetospheric pressure balance scenario.

\subsection{NGC 6397}
NGC 6397 has been intensively observed by optical, X-ray and radio instruments as a nearby GC with relatively \frcory{low} extinction ($D\approx 2.3~\mathrm{kpc}$; $E(B-V)=0.18$; \citealt[2010 edition]{harris1996}).  \citet{Ferraro01} identified an eclipsing MSP, PSR J1740-5340 (aka NGC 6397-A), the second-discovered ``redback" MSP, in NGC 6397. 
\citet{Grindlay01} used $49~\mathrm{ks}$ of {\it Chandra}/ACIS-I observation to reveal $25$ X-ray sources within $2\arcmin$ of the cluster, and used {\it Hubble Space Telescope} imaging to identify eight CVs and four ABs. \citet{Grindlay01} identified PSR J1740-5340 as an X-ray source (\frcory{U12}), and suggested that the similar (in X-ray and optical properties) source U18 might be a hidden MSP.
\citet{Bogdanov10} performed a much deeper search for X-ray sources in 
350 ks of {\it Chandra} observations, finding 79 sources. \citet{Cohn10} used new {\it Hubble} imaging to increase the totals for NGC 6397 to 15 CVs and 42 ABs. 

In this work, we present our detections of radio counterparts to U12, U18, two previously known faint X-ray sources (U97 and U108) and three newly detected \frcory{X-ray} sources (W127, W129 and W135). In Sec. \ref{sec:sec_obs}, we describe the observational data and relevant methodologies of data reduction; in Sec. \ref{sec:sec_results}, we present results from cross-matching X-ray with radio catalogues and discuss individual matches; in Sec. \ref{sec:sec_conclusion}, we summarise results and draw conclusions, and in Sec. \ref{sec:updated_X-ray_catalog}, we present an updated X-ray catalogue and tentative identifications of the new sources.

\section{Observations \& Analyses}
\label{sec:sec_obs}
\subsection{Radio Observations}
\label{sec:radio_obs}
NGC 6397 was observed by the Australia Telescope Compact Array (ATCA; PI: Strader) as part of the MAVERIC (Milky Way ATCA and VLA Exploration of Radio sources In Clusters) survey (Project Code: C2877; \citealt{Tremou18, Shishkovsky18}). The observation started on 2013-11-09 (exact times MJD 56605.94--56606.39, and 56606.96--56607.38) with two radio bands centered at $5.5$ and $9~\mathrm{GHz}$ (\frcory{both} with $2~\mathrm{GHz}$ of bandwidth) in the extended 6A configuration, for a total observational time of $20~\mathrm{hr}$ and a total on-source integration time of $16.2~\mathrm{hr}$. Calibration and image analysis was done with {\sc miriad} \citep{Sault95} and {\sc casa} \citep[version 4.2.0;][]{McMullin07}, rendering radio images at noise levels of $4.22$ and $4.81~\mujy~\mathrm{beam^{-1}}$. \frcory{We used the standard source PKS 1934-638 as both bandpass and flux calibrator, and the nearby source 1740-517 as our secondary calibrator, to determine the time-varying amplitude and phase gains that were then linearly interpolated to the target field.} \frcory{To achieve a high sensitivity, we applied the Briggs weighting scheme, with a robust parameter of 1, resulting in synthesized beamsizes of $1.54\arcsec \times 2.80\arcsec$ and $1.03\arcsec \times 1.87\arcsec$ at $5.5~\mathrm{GHz}$ and $9~\mathrm{GHz}$, respectively. Since there was not sufficient flux density in the field, no self-calibration was done.  
The radio positional accuracy directly obtained from this procedure is overestimated, so we inflate the source position uncertainties 
to at least 1/10 the $5.5/9~\mathrm{GHz}$ beamsizes.

We derive spectral indices ($\alpha$), defined by $S_\nu \propto \nu^\alpha$, using a Bayesian approach. We assume a flat prior of $\alpha$ between $-3.5$ and $1.5$ and calculate the posterior distribution for each radio source, from which we derive a median and an associated uncertainty range that covers $68\%$ of the total area. For sources that were not detected at $9~\mathrm{GHz}$ (so only upper limits to the fluxes are available), we report $3~\sigma$ upperlimits on $\alpha$ derived from the posterior distribution; medians are also calculated for these sources but are very sensitive to the prior, especially to the assumed lower bound ($-3.5$), so should be interpreted with caution.

More details on the procedure of generating $5~\sigma$ radio source catalogues will be presented in a separate work (Tudor et al., in prep).}

\subsection{X-ray Observations}
\label{sec:x_ray_obs}
We used the same {\it Chandra} dataset as in \citet{Bogdanov10}, including an ACIS-I observation from Cycle 1 (ObsID: 79; PI: Murray) and ACIS-S observations from Cycle 3 (ObsIDs: 2668, 2669; PI: Grindlay) and Cycle 8 (ObsIDs: 7460, 7461; PI: Grindlay). All level-1 files are first reprocessed to align to the most up-to-date calibrations using the {\tt chandra\_repro} script in the Chandra Interactive Analysis of Observations ({\sc ciao}; \citealt{Fruscione06})\footnote{\frcory{https://cxc.harvard.edu/ciao/}} software (version 4.11; CALDB 4.8.2). The resulting level-2 event files were used for further analyses. We then chose the longest observation (Obs. ID 7460) as the reference frame, to which we calculated relative offsets for all other observations based on the centroid positions of U17, the brightest source in the catalogue. These offsets were then used as input to the {\tt wcs\_update} tool to update aspect solutions for each observation. Finally, all offseted event files were combined using the {\tt merge\_obs} tool, rendering a combined event file. We then rebinned the combined event file to a quarter of an ACIS pixel ($0.25\arcsec$) and applied an energy filter of $0.5$-$7$ keV to generate an X-ray image. 

To get source positions, we run {\tt wavdetect} on a $400\arcsec \times 400\arcsec$ image centered on the cluster. We used size scale parameters of 1, 1.4, 2.0, and 2.8, while setting the threshold significance parameter to $3.9\times 10^{-7}$. This value is the reciprocal of the number of pixels in the image, \frcory{to} minimise misidentifications of background fluctuations as sources\footnote{\url{http://cxc.harvard.edu/ciao/threads/wavdetect/}}. The {\tt wavdetect} positions are then corrected for boresight offsets (Sec. \ref{sec:astrometry}) before being cross-matched with our radio catalogues. 

\begin{figure*}
    \centering
    \includegraphics[scale=0.25]{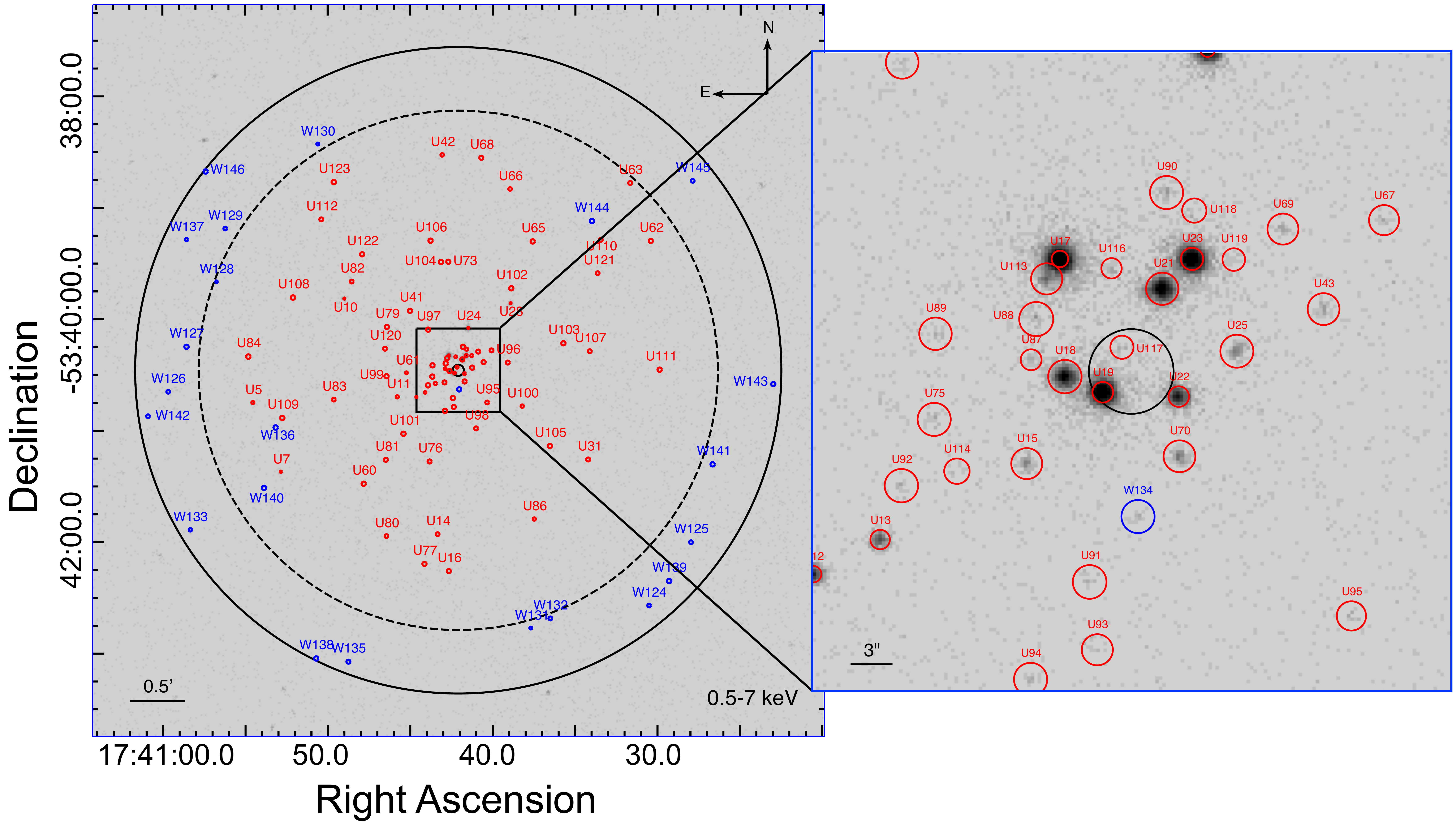}
    \caption{{\it Left:} $0.5$-$7$ keV X-ray image of a $6\farcm6 \times 6\farcm6$ square region centered on the cluster. North is up and east is to the left. New sources detected by {\tt wavdetect} (blue) and sources from \citet{Bogdanov10} (red) are marked by circles that enclose $90\%$ of their PSFs. The solid black circle shows the $2\farcm9$ half-light radius ($r_\mathrm{h}$), while the dashed black circle depicts the $2\farcm33$ searching radius used in \citet{Bogdanov10}. {\it Right:} a zoomed-in $45\arcsec \times 45\arcsec$ square region centered on the cluster. The solid black circle shows the $0.05\arcmin$ core radius ($r_\mathrm{c}$).}
    \label{fig:x_ray_img}
\end{figure*}

We searched for point sources within the half-light radius ($=2\farcm9$) derived by \citet[2010 edition]{harris1996} which is somewhat greater than that ($=2\farcm33$) used by \citet{Bogdanov10}. As a result, our {\tt wavdetect} run revealed a total of $23$ point sources \frcory{that are not previously catalogued}, of which $18$ are outside the $2\farcm33$ radius and $5$ are inside (Fig. \ref{fig:x_ray_img}). Some of these new sources are found to be positionally consistent with our radio sources (Sec. \ref{sec:sec_results}); moreover, X-ray positions derived from better astrometry with Gaia DR2 result in sub-arcsecond scale offsets (Sec. \ref{sec:astrometry}). We thus report an extended X-ray catalogue with updated source coordinates including both old and new sources (Sec. \ref{sec:updated_X-ray_catalog}). To distinguish from the old sources, the new sources are named with ``W" + sequential ID starting from 124, which are all summarised in Tab. \ref{tab:x_ray_catalog}. For each new X-ray source, we run the {\sc ciao} {\tt srcflux} script to calculate the X-ray counts in  soft ($0.5$-$1.5~\mathrm{keV}$) and hard ($1.5$-$6~\mathrm{keV}$) bands, and based on the total counts, we calculated 95\% error radii ($P_\mathrm{err}$) using an empirical formula from \citet{hong2005}.

For sources with radio counterparts, we also extracted X-ray spectra using the {\sc ciao} {\tt specextract} script and performed analyses using the {\sc heasoft/xspec} software (version 12.10.1; \citealt{Arnaud96})\footnote{\frcory{https://heasarc.gsfc.nasa.gov/xanadu/xspec/}}. We combined the ACIS-S spectra for each source using the {\sc heasoft/ftools} {\tt addspec} task to maximise spectral counts, and then rebinned the bright ($\gtrsim 700$ counts between $0.5$ and $6~\mathrm{keV}$) source spectra (U12, U18 and U24) to at least $20$ counts per energy bin, and faint spectra (U97, U108, W25, W129 and W135) to at least $1$ count per bin. The former are then modelled using the $\chi^2$ statistic, and the latter with the  C-statistic \citep{cash1979}. Since ACIS's sensitivity falls off at low energies, we only use energy channels between $0.5$-$10~\mathrm{keV}$ in all of our fits. We note that the spectra of U12 and U18 have been well-analysed by \citet{Bogdanov10}. 
We thus adopt the corresponding best-fitting models (power-laws for U12 and U18) and calculate X-ray fluxes between $1$ and $10~\mathrm{keV}$, which will be further used in Sec. \ref{sec:sec_results}. We fit the faint source spectra to \frcory{absorbed} power-law models using {\tt wilms} abundances \citep{Wilms00}, keeping the absorption column density ($N_\mathrm{H}$) fixed at the cluster value ($\approx 1.57\times 10^{21}~\mathrm{cm^{-2}}$, derived using $E(B-V)=0.18$ from \citealt{harris1996} and a \frcory{conversion} factor from \citealt{Bahramian15}). For W129, we obtained relatively more counts ($111$ counts between $0.5$ and $10~\mathrm{keV}$), so we fit its spectrum \frcory{allowing} $N_\mathrm{H}$ \frcory{to be free}.

\subsection{{\it HST} observations and Astrometry}
\label{sec:astrometry}
For absolute astrometry, we used data from observations by the {\it Hubble Space Telescope/Advanced Camera for Surveys} (HST/ACS) in the F625W band ($R_{625}$). The separate exposures are in the form of ``FLC" images that are pipe-lined, flat-fielded and cleaned for charge transfer efficiency (CTE) trails. We used the {\tt Tweakreg} and {\tt Astrodrizzle} tasks in the {\sc drizzlepac} software package to align and combine the individual FLC frames. We set the {\tt pixfrac} to $1.0$ and the final pixel size to $0.025\arcsec$, oversampling the resulting image by a factor of 2. 

We then aligned the resulting ``drizzle"-combined image to a catalogue with superior astrometry and used it as the reference frame. For this purpose, we chose stars that have relatively low astrometric uncertainties (with error in RA and DEC $\leq 0.05~\mathrm{mas}$) from the Gaia Catalogue of Data Release 2 \citep{gaia2016a, gaia2018}. These stars are matched with stars in the F625W frame so we can calculate averaged relative offsets. We found $50$ such stars within a searching radius of $1.2\arcmin$ centered on the cluster, resulting in average offsets (Gaia$-$ACS) of $=1.53\arcsec \pm 0.03\arcsec$ and $=1.28\arcsec \pm 0.01\arcsec$ in RA and DEC (1 $\sigma$ errors), respectively. We consider boresight correction for the catalogue by calculating relative offsets (ACS$-${\it Chandra}) between the X-ray centroids and positions of the identified counterparts for the three brightest CVs (U17, U19 and U23), from which we obtained average offsets in RA and DEC of  $=-0.09\arcsec$ and $-0.13\arcsec$, respectively. These offsets are applied to the combined {\it Chandra} image used for making finding charts. 

\begin{figure*}
    \centering
    \includegraphics[scale=0.35]{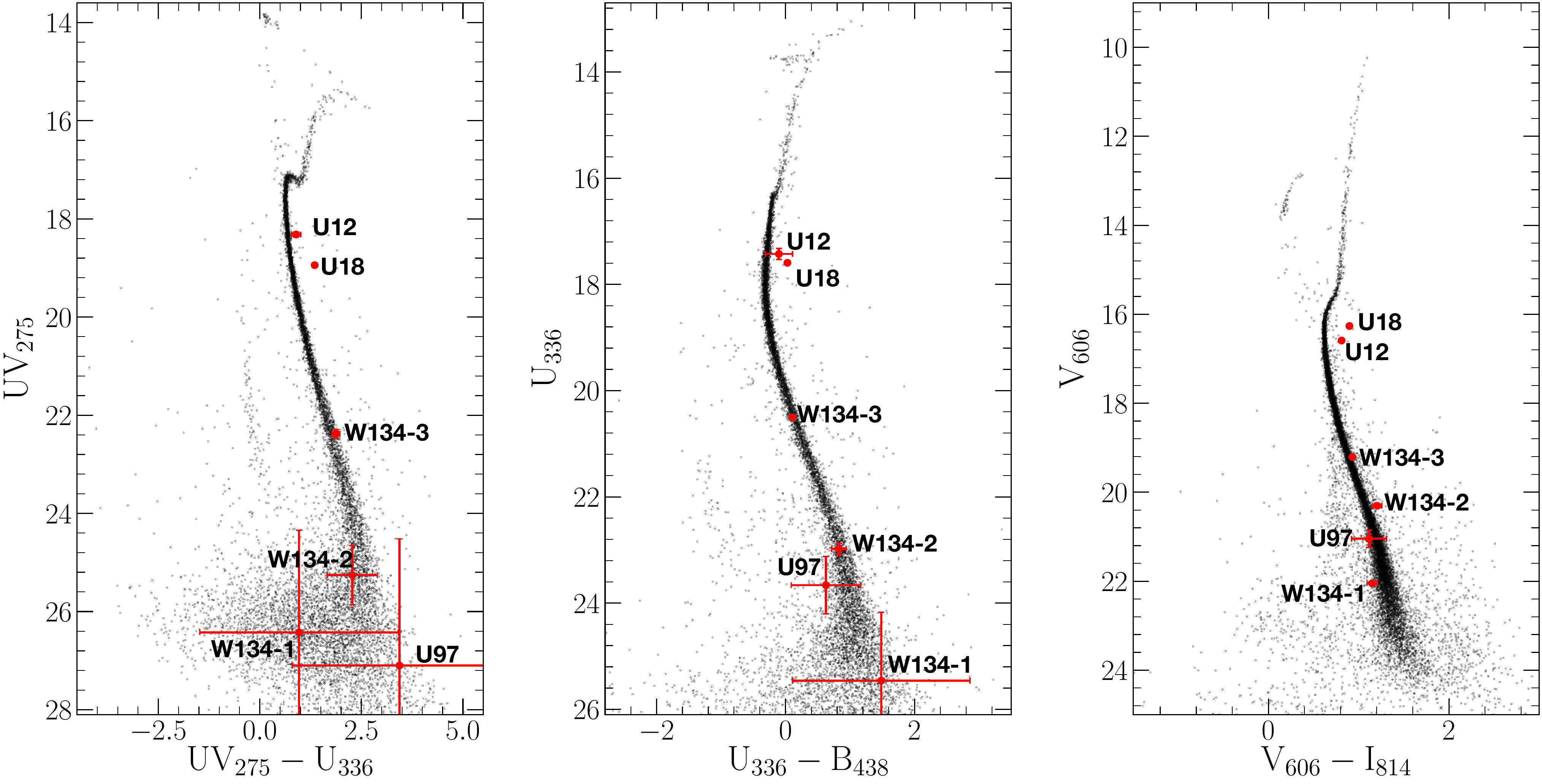}
    \caption{Colour-magnitude diagrams (CMDs) of stars from the HUGS catalogue. Locations and errors of \frcory{optical/UV} counterparts to X-ray sources with radio counterparts are indicated with red points and bars. We also present the photometry of $3$ objects that lie within the $95\%$ error circle of W134, a new X-ray source with HUGS data available (see also Appendix \ref{sec:updated_X-ray_catalog}).}
    \label{fig:hugs_cmds}
\end{figure*}

We also incorporate results from the {\it HST UV Globular Cluster Survey} (``HUGS"; \citealt{piotto2015, Nardiello18}) into our analyses, which provides photometry in three HST/WFC3 bands: F275W ($\mathrm{UV_{275}}$), F336W ($\mathrm{U_{336}}$), F438W ($\mathrm{B_{438}}$), and two HST/ACS bands: F606W ($\mathrm{V_{606}}$), F814W ($\mathrm{I_{814}}$). The latter are adapted from the {\it ACS Survey of Galactic Globular Clusters} (ACS GCS; GO-10775, PI: Sarajedini; \citealt{sarajedini2007, anderson2008}). We plotted 3 colour magnitude diagrams (CMDs) using stars that have good photometric measurements in multiple filters (Fig. \ref{fig:hugs_cmds}). Locations of interesting sources on the CMDs provide useful information in our further discussions (Sec. \ref{sec:sec_results}). 

\section{Results}
\label{sec:sec_results}
\begin{figure*}
    \centering
    \includegraphics[scale=0.35]{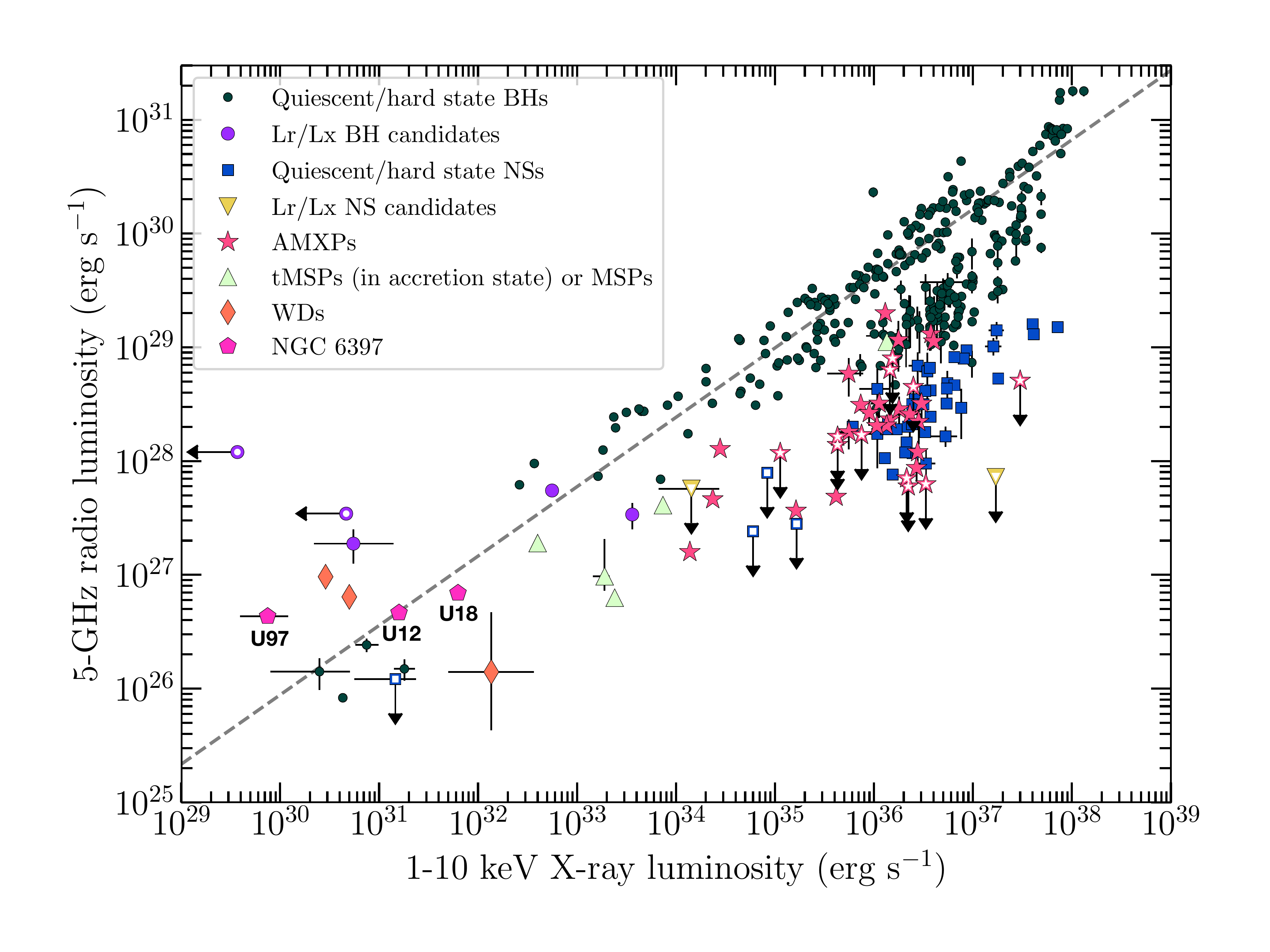}
    \caption{$5~\mathrm{GHz}$ radio luminosities plotted vs. $1$-$10~\mathrm{keV}$ X-ray luminosities for different classes of accreting compact objects. X-ray sources in NGC 6397 associated with radio counterparts are shown with filled pentagons. The radio luminosities are calculated assuming flat ($\alpha =0$) spectra. The deep-green circles represent known quiescent black holes \citep{Soleri10, MillerJones11, Gallo12, Ratti12, Corbel13, Rushton16, Plotkin17}. The dashed grey line shows the $L_R$-$L_X$ correlation for black holes from \citet{Gallo14}. The \frcory{larger} purple circles are radio-selected black hole candidates \citep{Strader12, Chomiuk13, MillerJones15, Tetarenko16, Bahramian17, Shishkovsky18}. The light-green triangles mark known transitional millisecond pulsars \citep{Hill11, Papitto13, Deller15, Bogdanov18}. The deep-blue squares and pink stars show quiescent/hard-state NSs and accreting millisecond X-ray pulsars (AMXPs), respectively \citep{Migliari06, Tudor17}. The yellow triangles show upper limits of $L_R$ of two X-ray transients \citep{Tetarenko16, Ludlam17}. The orange diamonds mark radio detected bright white dwarfs (WDs), including AE Aqr \citep{Eracleous91, Abada-Simon93}, SS Cyg (at flare peak; \citealt{Russell16}) and AR Sco (a radio-pulsating white dwarf; \citealt{Marsh16}). The hollow markers indicate upper limits. For script and data used to generate the plot, see \citet{Bahramian18}.}
    \label{fig:lrlx}
\end{figure*}

By cross-matching the $5~\sigma$ radio catalogue with our updated X-ray catalogue, we sought  radio sources that lie within $P_\mathrm{err}$, and found $7$ radio sources positionally \frcory{coincident} with {\it Chandra} sources, including U12 (PSR 1740-5340), U18, U97, U108 and $3$ new sources (W127, W129 and W135) that are likely AGNs. We found U24, the qLMXB, 
may be 
associated with a faint extension to a bright source in the $5.5~\mathrm{GHz}$ image, so we also include it in our discussions. In Tab.~\ref{tab:tab_positions}, we summarise the positional information of these sources, and in Fig.~\ref{fig:all_finders}, we show the corresponding X-ray, radio and/or optical finding charts. In Tab.~\ref{tab:tab_fluxes}, we present $1$-$10~\mathrm{keV}$ X-ray fluxes (from spectral fits described in Sec.~\ref{sec:x_ray_obs}), and in Fig.~\ref{fig:lrlx}, we plot the $5~\mathrm{GHz}$ radio vs. $1$-$10~\mathrm{keV}$ X-ray luminosities for cluster sources assuming flat ($\alpha=0$) radio spectra, together with other classes of accreting compact objects.

\begin{figure*}
    \centering
    \includegraphics[scale=0.31]{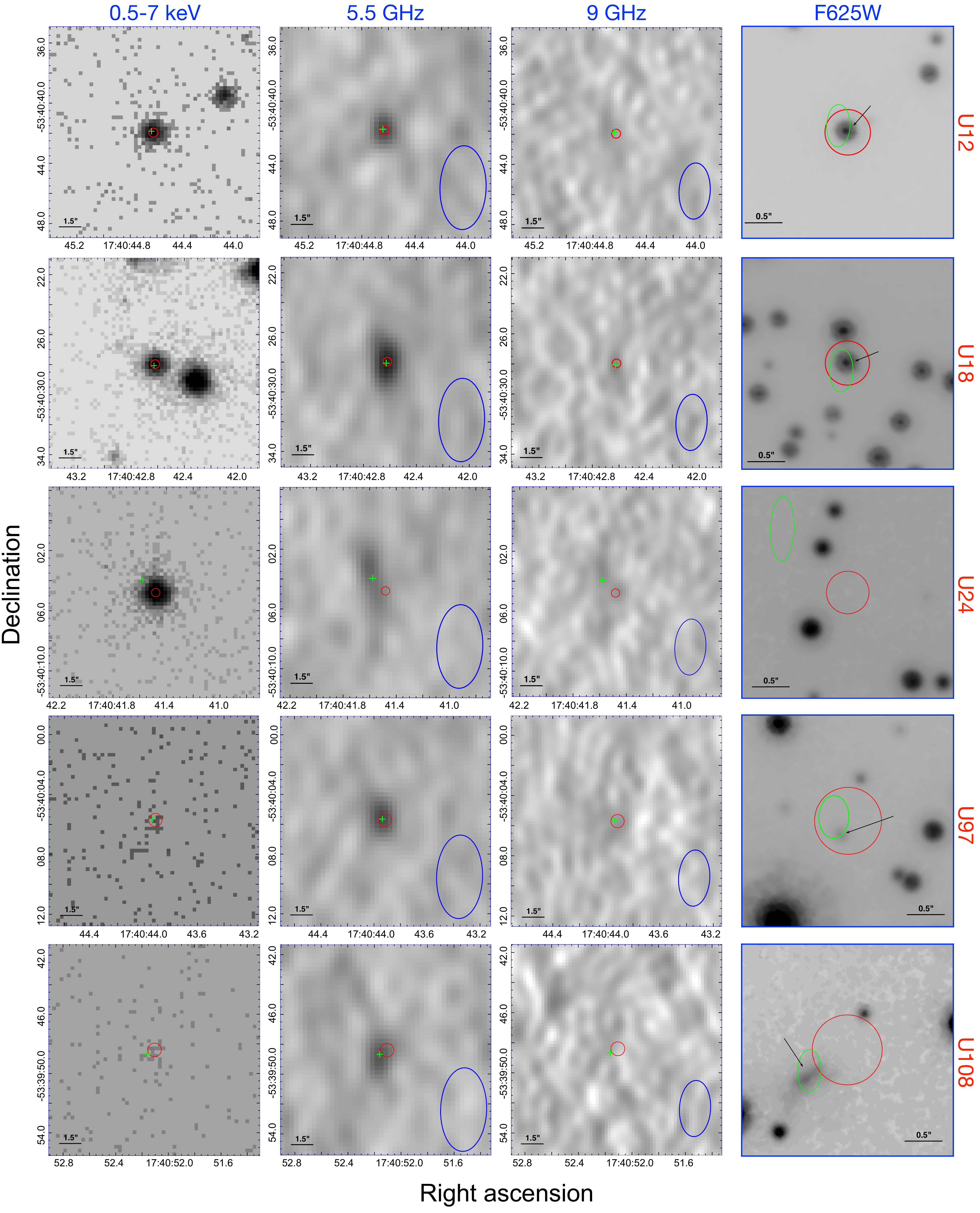}
    \caption{X-ray ($0.5$-$7~\mathrm{keV}$), radio ($5.5~\mathrm{GHz}$ and $9~\mathrm{GHz}$) and optical (F625W) finding charts for U12, U18, U24, U97 and U108. North is up and east is to the left. The X-ray and radio charts are all $14\arcsec\times 14\arcsec$ in size while the F625W finding charts are $2\farcs8 \times 2\farcs8$. We show the $95\%$ {\it Chandra} error region with red circles. Since the radio sources have relatively small positional uncertainties, we only show their nominal positions with green crosses in the X-ray and radio images, while in the somewhat zoomed-in F625W charts, we show radio error regions with green ellipses (the sizes of which are described in Sec. \ref{sec:sec_obs}) and indicate the optical counterparts with black arrows. The radio beams in the radio charts are shown with blue ellipses on the bottom right of the $5.5$ and $9~\mathrm{GHz}$ images.}
    \label{fig:all_finders}
\end{figure*}


\begin{figure*}
    \centering
    \includegraphics[scale=0.31]{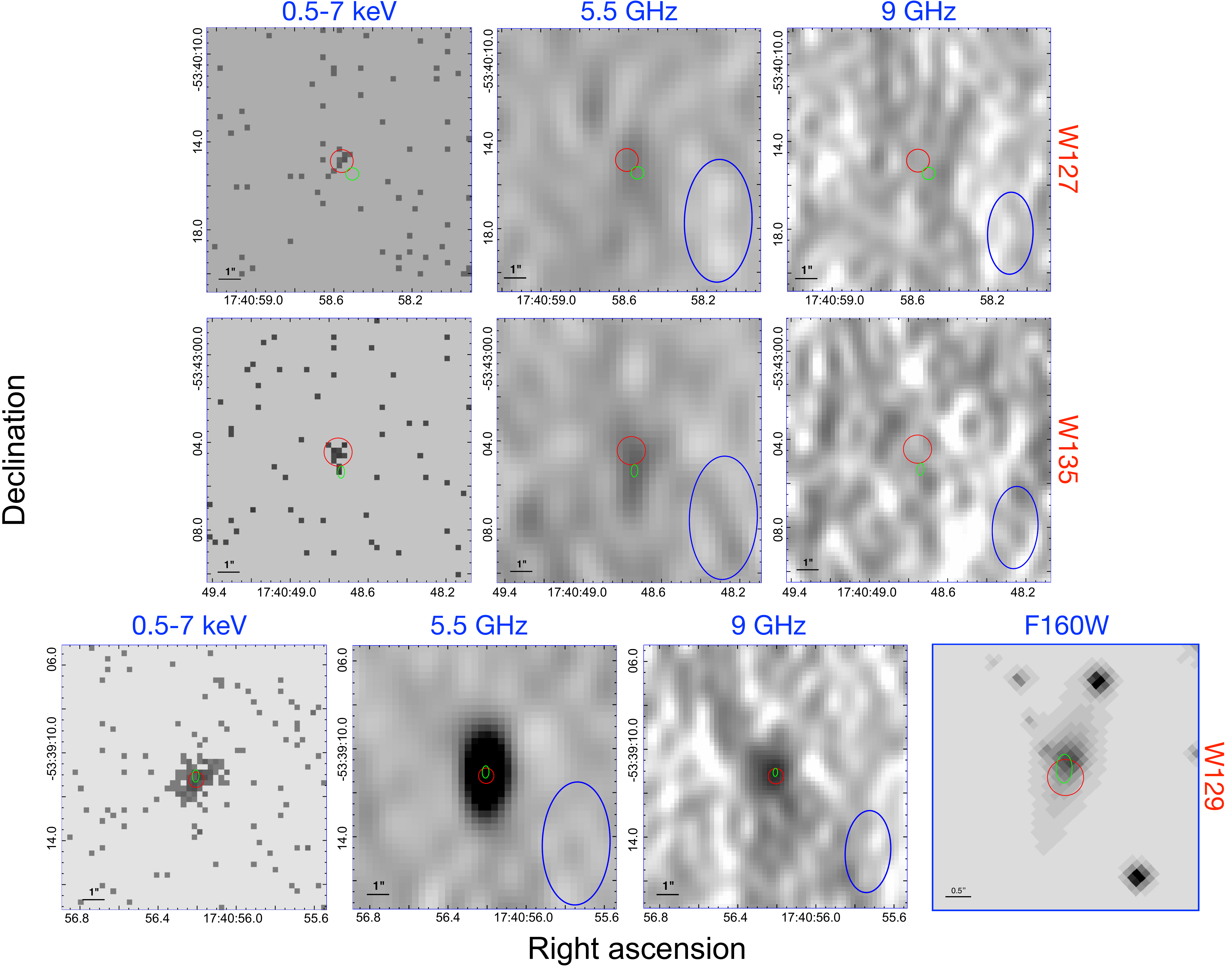}
    \contcaption{The top two rows show X-ray ($0.5$-$7~\mathrm{keV}$) and radio ($5.5~\mathrm{GHz}$ and $9~\mathrm{GHz}$) finding charts for W127 and W135. \frcory{The last row shows X-ray, radio and infrared (F160W) finding charts for W129. The $5.5$/$9~\mathrm{GHz}$ and the F160W charts are $12\arcsec \times 12\arcsec$ and $5\farcs5\times 5\farcs5$ in size, respectively. Radio error regions are shown with green ellipses.}}
\end{figure*}

\begin{table*}
    \centering
    \caption{Positional information of radio-X-ray cross-matched sources}
    \resizebox{\textwidth}{!}{
    \begin{tabular}{ccccccccc}
    \toprule
        Source ID & \multicolumn{2}{|c|}{X-ray position} & $P_\mathrm{err}^a$ & \multicolumn{2}{|c|}{Radio position$^b$} & \multicolumn{2}{|c|}{Optical position$^c$} & $P_\mu^d$ \\
                  & $\alpha_X$ & $\delta_X$ & $\arcsec$ & $\alpha_R$ & $\delta_R$ & $\alpha_O$ & $\delta_O$ & per cent \\
    \midrule
        U12  & 17:40:44.63 & $-$53:40:42.0 & 0.30 & 17:40:44.64(1) & $-$53:40:41.9(3) & 17:40:44.629 & $-$53:40:41.94 & $97.9$ \\
        U18  & 17:40:42.62 & $-$53:40:28.0 & 0.29 & 17:40:42.63(1) & $-$53:40:28.1(3) & 17:40:42.626 & $-$53:40:27.91 & $98.3$ \\
        U97  & 17:40:43.92 & $-$53:40:05.7 & 0.44 & 17:40:43.94(1) & $-$53:40:05.7(3) & 17:40:43.927 & $-$53:40:05.89 & $97.5$ \\
        U108 & 17:40:52.11 & $-$53:39:48.4 & 0.46 & 17:40:52.16(1) & $-$53:39:48.7(3) & 17:40:52.168 & $-$53:39:48.79 & - \\
        W127 & 17:40:58.56 & $-$53:40:14.9 & 0.52 & 17:40:58.51(2) & $-$53:40:15.5(3) & - & - & - \\
        W129 & 17:40:56.20 & $-$53:39:11.2 & 0.35 & 17:40:56.21(1) & $-$53:39:11.1(3) & 17:40:56.192 & $-$53:39:10.882 & -\\
        W135 & 17:40:48.75 & $-$53:43:04.4 & 0.64 & 17:40:48.74(1) & $-$53:43:05.4(3) & - & - & - \\
    \bottomrule
    \multicolumn{9}{l}{$^a$95\% error radii of X-ray positions; see \citet{hong2005}.}\\
    \multicolumn{9}{l}{\frcory{$^b$The uncertainties in the $\alpha_R$ and $\delta_R$ are derived from projections of the radio error ellipses (Sec. \ref{sec:sec_obs}) to the RA and DEC axes.}}\\
    \multicolumn{9}{l}{$^c$Uncertainties in optical positions are mostly from astrometry (Sec. \ref{sec:astrometry}).}\\
    \multicolumn{9}{l}{$^d$Membership probabilities from the HUGS catalogue; see \citet{Nardiello18}.}
    \end{tabular}
    }
    \label{tab:tab_positions}
\end{table*}

\begin{table*}
    \renewcommand{\arraystretch}{1.2}
    \centering
    \caption{Radio and X-ray properties of sources in Tab. \ref{tab:tab_positions}.}
    \begin{tabular}{cccccccc}
    \toprule
      ID & $F_X$ ($\times10^{-16}~\mathrm{erg~s^{-1}~cm^{-2}}$) & Photon index ($\Gamma$)$^a$ & \multicolumn{2}{c}{ $S_\nu$ ($\mathrm{\mu Jy}$)} & \multicolumn{2}{c}{Spectral index ($\alpha$)$^b$} & Comment \\
         &  $1$-$10~\mathrm{keV}$  & $F_X \propto E^{-\Gamma}$ & $5.5~\mathrm{GHz}$ & $9~\mathrm{GHz}$ & median & $3~\sigma$ upper limit & \\
    \midrule
       U12  & $250.4^{+14.8}_{-14.8}$ & $1.7^{+0.2}_{-0.2}$ & $36.7\pm 4.4$ & $<25.0$ & $-2.1^{+1.1}_{-1.0}$ & $<0.5$ & MSP \\
       U18  & $990.6^{+32.9}_{-33.0}$ & $1.3^{+0.1}_{-0.2}$ & $54.7\pm 4.3$ & $21.6\pm 4.4$ & $-2.0^{+0.4}_{-0.5}$ & - & MSP? \\
       U97  & $11.8^{+7.3}_{-5.6}$ & $0.5^{+1.0}_{-1.0}$ & $34.1\pm 4.7$ & $18.6\pm 4.5$ & $-1.3^{+0.6}_{-0.7}$ & - & AB or BH? \\
       U108 & $5.3^{+2.4}_{-1.9}$ & $1.9^{+0.8}_{-0.7}$ & $34.9\pm 4.5$ & $<26.8$ & $-2.0^{+1.2}_{-1.0}$ & $<0.7$ & Galaxy \\
       W127 & $7.2^{+4.6}_{-3.3}$ & $1.1^{+1.0}_{-1.0}$ & $26.9\pm 5.0$ & $<35.8$ & $-1.5^{+1.5}_{-1.3}$ & $<1.0$ & AGN? \\
       W129 & $416.8^{+72.3}_{-64.9}$ & $2.7^{+1.4}_{-1.2}$ & $163.0\pm 5.0$ & $65.0\pm 8.9$ & $-1.9^{+0.3}_{-0.3}$ & - & Galaxy \\
       W135 & $19.8^{+11.8}_{-8.7}$ & $-0.2^{+1.0}_{-1.2}$ & $31.5\pm 5.8$ & $<39.4$ & $-1.6^{+1.4}_{-1.3}$ & $<1.0$ & AGN? \\
    \bottomrule
    \multicolumn{8}{l}{$^a$X-ray photon indices derived from spectral fits to a power-law model (Sec. \ref{sec:x_ray_obs}).} \\
    \multicolumn{8}{l}{$^b$Radio spectral indices derived using a Bayesian approach; see Sec. \ref{sec:radio_obs}.}
    \end{tabular}
    \label{tab:tab_fluxes}
\end{table*}

\subsection{U12, PSR J1740-5340}
PSR J1740-5340 was discovered by the Parkes telescope as part of a $1.4~\mathrm{GHz}$ radio timing survey for GC MSPs \citep{DAmico01a, DAmico01b}. Irregular radio eclipses were found for more than $40\%$ of its $\sim 33$-hr orbit. 
\frcory{The X-rays also show a decrease during the radio eclipse, which was }
interpreted as occultations by the companion star of the shock front produced by interaction between pulsar winds and outflowing mass from the companion \citep{Bogdanov10}. In the optical, a sub-subgiant with $M\sim 0.3 M_\odot$ was 
\frcory{identified as the }
counterpart to this MSP \citep{Ferraro01, Kaluzny03,orosz2003}. Further optical spectroscopic studies of this sub-subgiant revealed asymmetric line features and ellipsoidal photometric variations, indicating 
that the companion fills its Roche lobe \citep{orosz2003}, and that it is \frcory{heated} by a beam of radiation from the pulsar \citep{Ferraro03, Sabbi03}.

Consistently, we found the optical counterpart to U12 appears to be in the sub-subgiant region on all three HUGS CMDs (Fig. \ref{fig:hugs_cmds}). 
There is no sign of strong UV emission, indicating the absence of a hot disc. This is consistent with the picture suggested by \citet{Bogdanov10} where matter flows from the companion towards the NS, never reaches the NS, but rather is halted and swept back by the strong pulsar wind, forming an intrabinary shock.
\\

We found an $8~\sigma$ ATCA source $0.15\arcsec~(\approx 0.5 P_\mathrm{err})$ from the nominal X-ray position of U12 (Fig. \ref{fig:all_finders}), with a $5.5~\mathrm{GHz}$ flux density of $36.7 \pm 4.4~\mujy$. The $9~\mathrm{GHz}$ observation reveals an upper limit of $25~\mujy$, so we obtain 
a rough $3~\sigma$ constraint on the spectral index of $<0.5$ (see Sec. \ref{sec:radio_obs}). 

We then imaged the ATCA 5.5 GHz data in two \frcory{(similar-length)} parts, inside and outside the eclipse phases. To calculate the eclipse phases, we used the eclipse phases of 0.05-0.45 quoted in \citet{DAmico01a}, the orbital period of 1.35405939(5) days from \citet{DAmico01b}, and a T0 from \citet{Mucciarelli13} of MJD 52413.22761, to give relevant eclipse \frcory{times} of MJD 56605.44-56605.98 and 56606.79-56607.33.
The out-of-eclipse image detects U12 at $58.3 \pm 5.1~\mujy$, while the in-eclipse image does not detect U12, with a $3~\sigma$ upper limit of $19~\mujy$. 

Comparing the out-of-eclipse flux density to the $1.4~\mathrm{GHz}$ flux density of $S_\nu \sim 0.5-1.5~\mathrm{mJy}$ measured by the Parkes radio telescope \citep[][\frcory{we assume this value is outside the eclipse only}]{DAmico01a} constrains the spectral index  $-2.4 \lesssim \alpha \lesssim -1.6$, within the range of steep spectral indices observed in radio pulsars \citep{Bates13}. We also created images in Q, U, and V Stokes parameters, but did not see evidence of polarised emission. The $3~\sigma$ upper limit of  $16.5~\mujy$/beam indicates a polarisation upper limit of $28\%$. 

\subsection{U18: A hidden MSP?}
\citet{Grindlay01} discovered the X-ray source U18, associated it with a sub-subgiant counterpart \citep[see also][]{Cohn10}, and suggested, given the similarities in X-ray and optical properties to PSR J1740-5340 (U12), that U18 is also a redback MSP, where, in this case, the eclipsing wind completely (or near-completely) blocks the observability of radio pulsations. \cite{Bogdanov10} confirmed U18's  hard non-thermal spectrum ($\Gamma\approx 1.3$) and variability on timescales from hours to years, which are consistent with a shocked-wind MSP origin for the X-ray emission. \citet{Cohn10} confirmed that U18's proper motion marked it as a cluster member, and identified strong optical variability from U18 ($\sigma\sim0.03$ mags).

U18 has a definite match with an $11.7~\sigma$ radio source 
$0.13\arcsec~(\approx 0.4P_\mathrm{err})$ from the nominal X-ray position in our $5.5~\mathrm{GHz}$ radio image (Fig. \ref{fig:all_finders}). The source was also detected in $9~\mathrm{GHz}$, 
allowing us to \frcory{measure} its spectral index, $\alpha = -2.0^{+0.4}_{-0.5}$. The steep radio spectrum is consistent with observations of many MSPs \citep{Sieber73, Lorimer95, Bates13}. 
Breaking the $5.5~\mathrm{GHz}$ data into two days \frcory{of similar exposures} gives flux densities of $67 \pm 5$ and $40 \pm 6$ $\mujy$, confirming variability (\frcory{at $3.5~\sigma$ significance}), but not \frcory{indicating} extreme 
flaring. 
Searching for polarisation gave a $3~\sigma$ upper limit of $25\%$ polarised emission (similar to U12 above). 
We also confirm the sub-subgiant position of U18 in all three CMDs, including the UV HUGS data (Fig. \ref{fig:hugs_cmds}).

Could U18 be something other than an MSP? The radio and X-ray luminosities of U18 are compatible with quiescent black holes \citep[e.g.][]{Gallo14}. However, the steep radio spectral index of U18 \frcory{strongly disfavours this scenario, in that quiescent BHs generally have flat to inverted radio spectra}.

U18's steep radio spectral index and high radio luminosity \frcory{($L_R\approx 6.9\times 10^{26}~\mathrm{erg~s^{-1}}$ at $5.5~\mathrm{GHz}$, if one incorrectly assumes a flat radio spectrum)} without short, bright flares set its radio emission apart from the CV behaviour discussed in Sec. \ref{sec:gc_radio_src}. Nor do the unusual systems AE Aqr (a propeller system) or AR Sco (a WD pulsar) match U18's spectral index, or AE Aqr's flaring behaviour. 
The optical counterpart to U18 exhibits a moderate $\mathrm{H}\alpha$ excess, and a relatively low X-ray/optical ratio, which are both consistent with typical globular cluster CVs \citep[e.g.][]{Cohn10}. 
However, U18 completely lacks (Fig. \ref{fig:hugs_cmds}) the UV or blue excesses that originate from the hot accretion disc and/or WD surface in typical CVs. Marked optical/UV variabilities are common in some CVs, which might alter UV \frcory{flux by up to $1$ magnitude} on timescales of hours (see e.g., \citealt{RiveraSandoval18}), but cannot be responsible for the strong red excesses of U18  observed in all 3 CMDs. Considering the radio and UV/optical data, we rule out a CV interpretation for U18. 

Finally, an AB interpretation cannot be completely ruled out (as the radio spectral index is reasonable; \citealt{Garcia03}). \cory{The radio luminosity ($L_R =6.9\times10^{26}~\mathrm{erg~s^{-1}}$) can be reached by strong flares in some RS CVn systems \citep[see e.g.,][]{Garcia03}, but is far above the steady radio luminosities seen for known low-mass ABs \citep[see e.g.,][]{Guedel93}.}

Thus the evidence points to a hidden MSP in U18. Nondetection of pulsations in 1.4 and 3 GHz searches,\footnote{Although not published, D'Amico et al. have been timing PSR J1740-5340 at Parkes for a decade at 3 GHz, \frcory{and their beam includes U18}.} if confirmed, generally could imply an unfavorable beaming geometry, but the detection of a bright, steep-spectrum $5.5~\mathrm{GHz}$ source is strong evidence that we are seeing radio pulsar emission from U18. It is possible that U18 shows pulsations at 5.5 GHz while being completely eclipsed at 3 GHz, so we do encourage higher-frequency pulsar searches of NGC 6397. However, this seems unlikely; we conclude that U18 is an MSP whose pulsations 
\frcory{may have been} completely eclipsed in all observations and frequencies used so far. We discuss the eclipse mechanism further in Sec. \ref{sec:sec_discussion}.

\subsection{U24--radio related to a quiescent LMXB?}
U24 is a quiescent low-mass X-ray binary containing a neutron star \citep{grindlay2001,Guillot11,Heinke14}. Detection of radio emission from such an object would be unprecedented.
A $6~\sigma$ \frcory{flat-to-steep spectrum} ($\alpha = -0.7^{+0.6}_{-0.6}$) ATCA source was found $1.2\arcsec$ ($\approx 4.3P_\mathrm{err}$) northeast of the X-ray position. The relative astrometry indicates that this radio source is not associated with U24, nor with any other detectable X-ray or optical source. However, the radio source appears extended, and may have a faint ($\lesssim 3\sigma$) feature $\sim 1.5\arcsec$ south of the radio centroid which may overlap with the position of U24 (displaced by $\approx 0.2\arcsec$,
Fig.~\ref{fig:all_finders}). \frcory{This faint extension could be a separate source, though the data is too low-significance to be sure.} Deeper \frcory{radio} imaging of NGC 6397 would be needed to verify whether we indeed see radio emission from this quiescent neutron star. 



\subsection{U97}
We found a $7.3~\sigma$ ATCA source $0.13\arcsec~(0.27P_\mathrm{err})$ from the nominal X-ray position of U97, which shows a spectral index of intermediate steepness ($\alpha=-1.3^{+0.6}_{-0.7}$). \frcory{Assuming the cluster distance and a flat spectrum, we estimate a $5.5~\mathrm{GHz}$ radio luminosity of $4.3\times 10^{26}~\mathrm{erg~s^{-1}}$.}

\citet{Cohn10} found a faint ($V \approx (B + R)/2 \approx 21.9$) and moderately blue counterpart to U97, exhibiting prominent H$\alpha$ variability. Such variation is rare, so the suggested counterpart is highly likely to be the true counterpart.
U97 is located fairly close ($\approx 0.5\arcmin$ or $\approx 0.2r_\mathrm{h}$) to the cluster centre, and the suggested counterpart's proper motion matches the cluster, indicating a high membership probability  (\citealt{Nardiello18}; Tab. \ref{tab:tab_positions}), so U97 is very likely a cluster member.


\subsubsection{A chromospherically active binary?}
The X-ray/optical ratio of this source overlaps with those of cluster ABs. The HUGS photometry yields a somewhat brighter $\mathrm{V_{606}}$ magnitude of $21.0\pm 0.2$, 
suggesting \frcory{significant variability.} 
\frcory{The observed $V$ band magnitude converts to an absolute magnitude ($M_V$) of $9.2$, corresponding to a lower main sequence star of roughly $0.05L_\odot$ and $M\approx 0.54 M_\odot$, according to the stellar models of \citet{pecaut2013}. This suggests that U97 may be a BY Draconis variable, that is, an AB composed of two MS stars.}



\citet{Guedel93} found that $\log(L_X/L_R) \lesssim 15.5$ for multiple types of ABs ($\approx$ applies for less luminous objects like BY Dra, while $<$ applies for more luminous classes such as RS CVn), suggesting that the heating of hot coronae (which emit X-rays) and the acceleration of particles (which emit in the radio) are closely related. Given the observed $L_X$ ($\approx 7\times10^{29}~\mathrm{erg~s^{-1}}$, see Tab. \ref{tab:tab_fluxes}) this correlation would 
\frcory{predict} 
$L_R(8.5~\mathrm{GHz})\approx 2.4\times 10^{14}~\mathrm{erg~s^{-1}~Hz^{-1}}$, which is $\sim 3$ orders of magnitude fainter than our observed $9~\mathrm{GHz}$ luminosity. The X-ray and radio observations were not simultaneous, so \frcory{conceivably} the high radio luminosity could have been caused by flares at the time of the ATCA observations, while {\it Chandra} may have observed the source in a quiescent state. \frcory{However, we found no clear sign of variability from comparing the two separate ATCA images.}
 \frcory{ Furthermore, the radio luminosity ($L_R(5.5~\mathrm{GHz})\approx 4.3\times 10^{26}~\mathrm{erg~s^{-1}}$) we observe} would be unparalleled for BY Dra systems. This scenario must be judged extremely unlikely.




\subsubsection{A quiescent black hole (BH) X-ray binary}
The major argument in favour of a BH scenario is that the $L_R/L_X$ ratio is consistent with the scatter of the $L_R-L_X$ correlation for quiescent stellar-mass BHs, from \citet{Gallo14}. The radio spectral index is not well constrained ($\alpha = -1.3^{+0.6}_{-0.7}$) but could be marginally consistent with the flat-to-inverted spectrum observed in quiescent systems (see e.g., \citealt{Gallo19}), interpreted as self-absorbed synchrotron emission from a collimated jet.
The low X-ray luminosity ($L_X (1-10~\mathrm{keV}) = 7.5^{+4.6}_{-3.5}\times 10^{29}~\mathrm{erg~s^{-1}}$; Tab. \ref{tab:tab_fluxes}) is also consistent with a BH nature.


Although the fit quality is limited by the dearth of counts ($27$ counts between $0.5$ and $10~\mathrm{keV}$), the (moderately) hard photon index ($\Gamma = 0.5\pm 1.0$) of U97 could be a counterargument to this scenario. X-ray spectra of quiescent BHs 
are usually characterised by softer power-laws ($\Gamma\sim 2.1$; \citealt{Tomsick01, Kong02, Corbel06, Plotkin13, Reynolds14}). \frcory{Conceivably a quiescent BH, seen edge-on, would present a harder X-ray spectrum, as the X-rays could be heavily absorbed by part of the accretion disc.}

Although the optical counterpart is not obviously blue (as might be expected in an accretion disk scenario), the quiescent nature of the disc at this time could permit the optical/UV light to be dominated by the companion star \frcory{(e.g., in XTE J1118+480 and A0620-00; see \citealt{Gallo07})}.



\subsubsection{A white dwarf system?}
As mentioned previously in Sec. \ref{sec:gc_radio_src}, CVs are often observed to be radio sources, but they typically \frcory{have rather low luminosities ($\lesssim 10^{25}~\mathrm{erg~s^{-1}}$)} in the radio \citep[e.g.][]{Barrett17}. Two highly magnetic white dwarfs are unusually bright in the radio; AE Aqr \citep{Eracleous91} and AR Sco \citep{Marsh16}. Both lie quite near to U97 in the X-ray/radio luminosity diagram (Fig. \ref{fig:lrlx}). Note that neither is a standard, accreting CV; AE Aqr is largely in the propeller regime \citep{Eracleous96}, while AR Sco is not thought to be accreting, but to be the first known white dwarf pulsar. 
The negative radio spectral index of U97 contrasts with the positive radio spectral index of AR Sco, while AE Aqr attains its maximum radio flux density only in short flares, while we measure U97 to have the same radio flux densities (37 and 38 $\mujy$, respectively, with noise of 5.5/6.8 $\mujy$/beam respectively) in the two halves of our observation. Both AR Sco and AE Aqr are also quite blue, especially in the ultraviolet. This  \frcory{appears not to be the case for U97, according to the HUGS photometry (Fig. \ref{fig:hugs_cmds})}.


\subsection{U108, W127, W129 and W135}
\label{sec:sec_likely_agns}
U108 is near an extended object on the optical images, which was interpreted as  interacting galaxies by \citet{Cohn10}. We found a $7.7~\sigma$ radio source $0.57\arcsec$ ($1.2P_\mathrm{err}$) from the nominal X-ray position, slightly outside the 95\% error circle; however, the radio position agrees closely with the optical counterpart, suggesting that the radio and X-ray sources are probably associated with these galaxies (Fig. \ref{fig:all_finders}).

W127, W129 and W135 are also likely to be extragalactic sources. \cory{Arguments for their non-member nature include their large offsets from the cluster centre and, more importantly, their \frcory{apparently} high \frcory{X-ray} absorption \frcory{(see below)},
common in AGNs from the obscuring torus. 
W129 has a relatively ``soft" photon index of $\Gamma=2.7^{+1.4}_{-1.2}$ (Tab. \ref{tab:tab_fluxes}), but \frcory{this fit requires heavy absorption of}  $N_\mathrm{H}=1.2^{+6.6}_{-4.7}\times 10^{23}~\mathrm{cm^{-2}}$ (see Fig. \ref{fig:spec_w35_w125}). \frcory{In fact, we searched in the {\it Mikulski Archive for Space Telescopes} (MAST)\footnote{http://archive.stsci.edu/} and found that the position of W129 is marginally covered by an {\it HST} WFC3/IR observation, where we visually identified that W129 is associated with an extended source, likely an early-type galaxy (Fig. \ref{fig:all_finders}).} W127 is deficient of soft X-ray counts, so seems also to be affected by absorption (see Tab.\ref{tab:x_ray_catalog}). A power-law fit to W135 results in a very hard $\Gamma=-0.2^{+1.0}_{-1.2}$. A negative value in this case fits the soft part of the spectrum, since we fixed the $N_\mathrm{H}$ to the cluster value, which is too low to compensate for the dearth of soft photons. Moreover, searches for possible optical counterparts to either W127 or W135 in the {\it Gaia} DR2 database did not reveal any objects up to twice the X-ray error radii (Sec. \ref{sec:updated_X-ray_catalog}), suggesting an extragalactic nature.
}


\section{Discussion}
\label{sec:sec_discussion}
The identification of U18 as a hidden MSP via its steep-spectrum radio emission is extremely exciting. Its identification in the second-nearest globular cluster (making U18 one of the three nearest MSPs in globular clusters known) suggests that a substantial population of similar systems may exist, with implications for our understanding of the evolution of X-ray binaries into MSPs. 

Evidence from previous studies (see Sec. \ref{sec:sec_eclipsing_msps}) suggests that more than one mechanism for pulsar eclipses operates, with several works showing that the pulsar signal is absorbed at low frequencies ($<$1 GHz), but may be scattered at high frequencies. Our current study is unique in \frcory{indicating} that two different eclipse mechanisms \frcory{appear to be} operating in two different MSPs at 5.5 GHz. U12 suffers absorption (dimming by at least a factor of 2.8) during its $5.5~\mathrm{GHz}$ eclipse, while U18 has a similar radio flux as U12 but remains undetected \frcory{as a radio pulsar}. \frcory{Our detection of the  radio counterpart to} U18 strongly indicates a scattering process is at work in this system. It is particularly odd that U12 and U18 would show different eclipse mechanisms, as the two systems are remarkably similar in their radio flux, their X-ray flux, and their companion properties (both are sub-subgiants, located a magnitude below the subgiant branch). 
 All three HUGS CMDs show a greater red excess in U18 than U12 (Fig. \ref{fig:hugs_cmds}). This may suggest that the NS in U18 is in a wider orbit, so that less mass has been stripped away from its companion, leaving it redder \citep{Ivanova17}. How this could relate to different absorption mechanisms and properties is unclear.

As suggested by \citet{Tavani91}, it may be possible to search for gamma-ray pulsations from U18 to confirm its nature. Blind gamma-ray pulsation searches are only effective for isolated MSPs \citep{Clark18}, but the example of PSR J1311-3430 \citep{Romani12,Pletsch12} shows that gamma-ray pulsations can be uncovered if an optical orbital ephemeris for the counterpart is available. Indeed, the orbital solution for U18 has recently been uncovered by Pichardo Marcano, Rivera Sandoval, et al. (in prep.), suggesting the feasibility of such a search.

\section{Conclusions}
\label{sec:sec_conclusion}
Our studies of the {\it Chandra} and ATCA observations reveal radio continuum counterparts to $7$ X-ray sources including a known MSP (U12; PSR 1740-5340), a ``hidden" pulsar (U18) likely obscured by matter stripped from the companion, a BH candidate (U97), a previously identified extragalactic source (U108) and three new sources (W127, W129 and W135), of which two (W127 and W135) are likely to be extragalactic, while the other, W129 has a definite match with a galaxy. The similarities between U12 and U18 and the detection of a steep-spectrum radio counterpart to U18 indicate different eclipsing mechanisms in these two sources, and we suggest that scattering is likely to be the dominant mechanism at high radio frequencies. 

\section*{Acknowledgements}
COH is supported by NSERC Discovery Grant RGPIN-2016-04602, and a Discovery Accelerator Supplement. J.S. acknowledges support from NSF grant AST-1308124 and a Packard Fellowship. JCAM-J is the recipient of an Australian Research Council Future Fellowship (FT140101082), funded by the Australian government. L.C. acknowledges support from Chandra G06-17040X. This work is primarily based on observations obtained with {\it Chandra} and {\it ATCA}. {\it Chandra} data reduction and analyses made use of the software provided by the Chandra X-ray Center (CXC) in the application package {\sc CIAO}. X-ray spectral analyses in this work made use of software packages provided by the High Energy Astrophysics Science Archive Research Center (HEASARC), which is a service of the Astrophysics Science Division at NASA/GSFC and the High Energy Astrophysics Division of the Smithsonian Astrophysical Observatory. The Australia Telescope Compact Array is part of the Australia Telescope National Facility which is funded by the Australian Government for operation as a National Facility managed by CSIRO. We also used data from the European Space Agency (ESA) mission {\it Gaia} (\url{https://www.cosmos.esa.int/gaia}), processed by the {\it Gaia} Data Processing and Analysis Consortium (DPAC, \url{https://www.cosmos.esa.int/web/gaia/dpac/consortium}). Funding for the DPAC has been provided by national institutions, in particular the institutions participating in the {\it Gaia} Multilateral Agreement. 




\bibliographystyle{mnras}
\bibliography{ref} 




\appendix

\section{An updated X-ray source catalogue and tentative identification of new sources}
\label{sec:updated_X-ray_catalog}

\begin{figure*}
    \centering
    \includegraphics[scale=0.5]{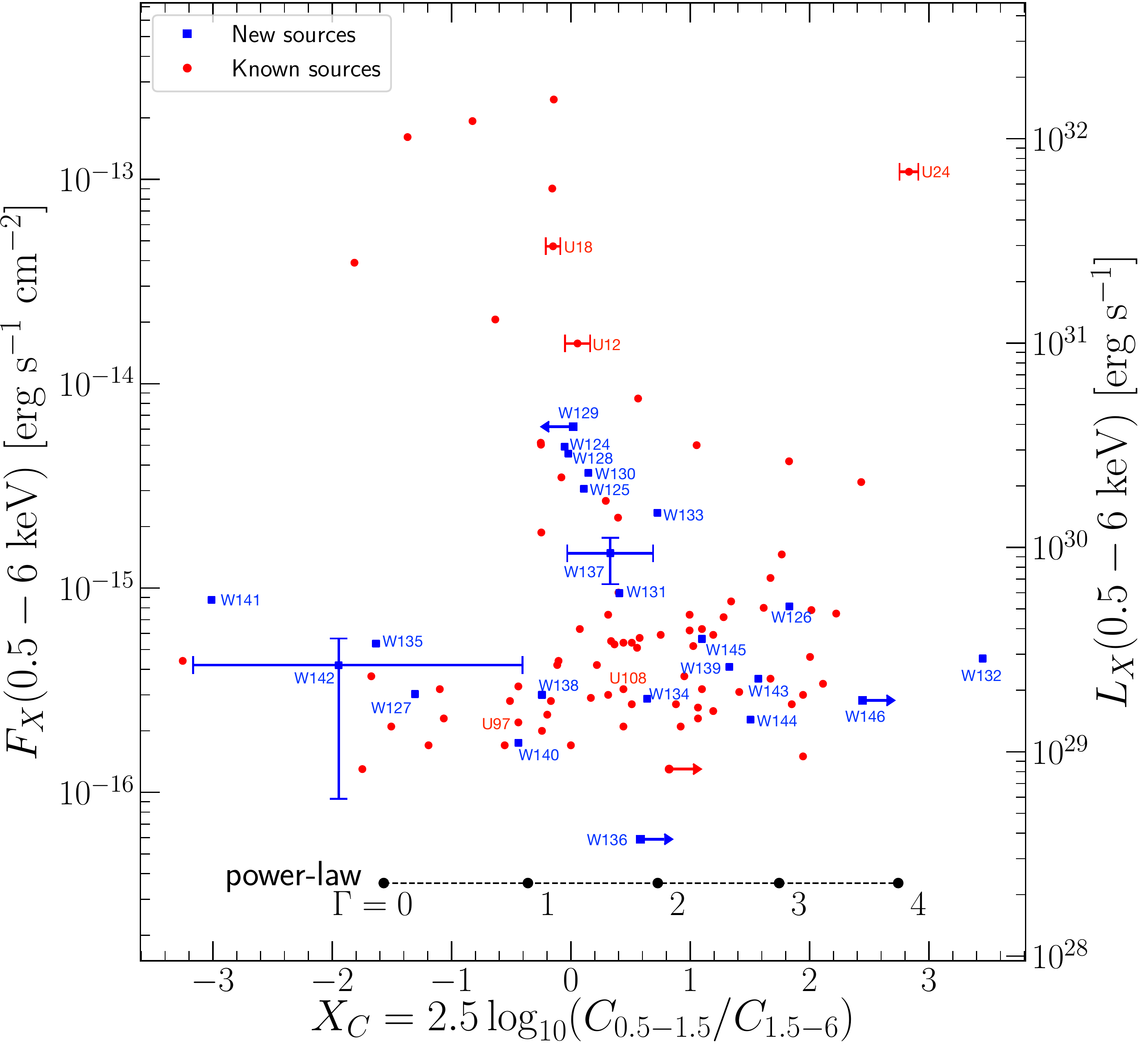}
    \caption{X-ray colour-magnitude diagram plotting $0.5$-$6~\mathrm{keV}$ fluxes (left scale) and luminosities (right scale) vs. hardness ratio ($X_C$, defined in eq.(\ref{eq:X_c})) for known (red) and new (blue) sources. $X_C$s and fluxes of the known sources are adapted from \citet{Bogdanov10}, while were calculated by {\tt srcflux} (Sec. \ref{sec:x_ray_obs}) for the new sources. For better readability, we only labelled the relevant known sources and the new sources, and put error bars on several sources to represent uncertainties at different flux levels. Upper/lower limits and errors in $X_C$ are at the 90\% confidence as derived according to methods in \citet{gehrels1986}. We also indicate the locations of power-law models with different photon indices ($\Gamma$s).}
    \label{fig:xray_cmd}
\end{figure*}

Our X-ray catalogue (Tab. \ref{tab:x_ray_catalog}) includes sources within the $2\farcm9$ half-light radius ($r_\mathrm{h}$, as updated by \citealt{harris1996}, 2010 revision; note that \citealt{Bogdanov10} used a previous Harris catalog version with  $r_\mathrm{h}=2.33\arcmin $ to search for sources) and source positions with updated astrometry from {\it Gaia} DR2. We adopt source counts from \citet{Bogdanov10} and calculate counts in the same bands using the {\sc ciao} {\tt srcflux} tool, for the new sources (Sec. \ref{sec:x_ray_obs}). Most of these sources are outside the {\it HST} field of view, so direct photometric identification might be hard. However, tenuous classification is still possible solely based on the X-ray hardness. For this purpose, we define an X-ray hardness ratio

\begin{equation}
    X_C \equiv 2.5 \log_\mathrm{10}\left( \frac{C_\mathrm{0.5-1.5~keV}}{C_\mathrm{1.5-6~keV}} \right),
    \label{eq:X_c}
\end{equation}
where $C_\mathrm{0.5-1.5~keV}$ and $C_\mathrm{1.5-6~keV}$ are X-ray counts in $0.5$-$1.5~\mathrm{keV}$ and $1.5$-$6~\mathrm{keV}$, respectively. The distribution of $X_C$ is shown in Fig. \ref{fig:xray_cmd}, which plots $0.5$-$6~\mathrm{keV}$ fluxes vs. $X_C$ for known and new sources.

\begin{figure}
    \centering
    \includegraphics[scale=0.48]{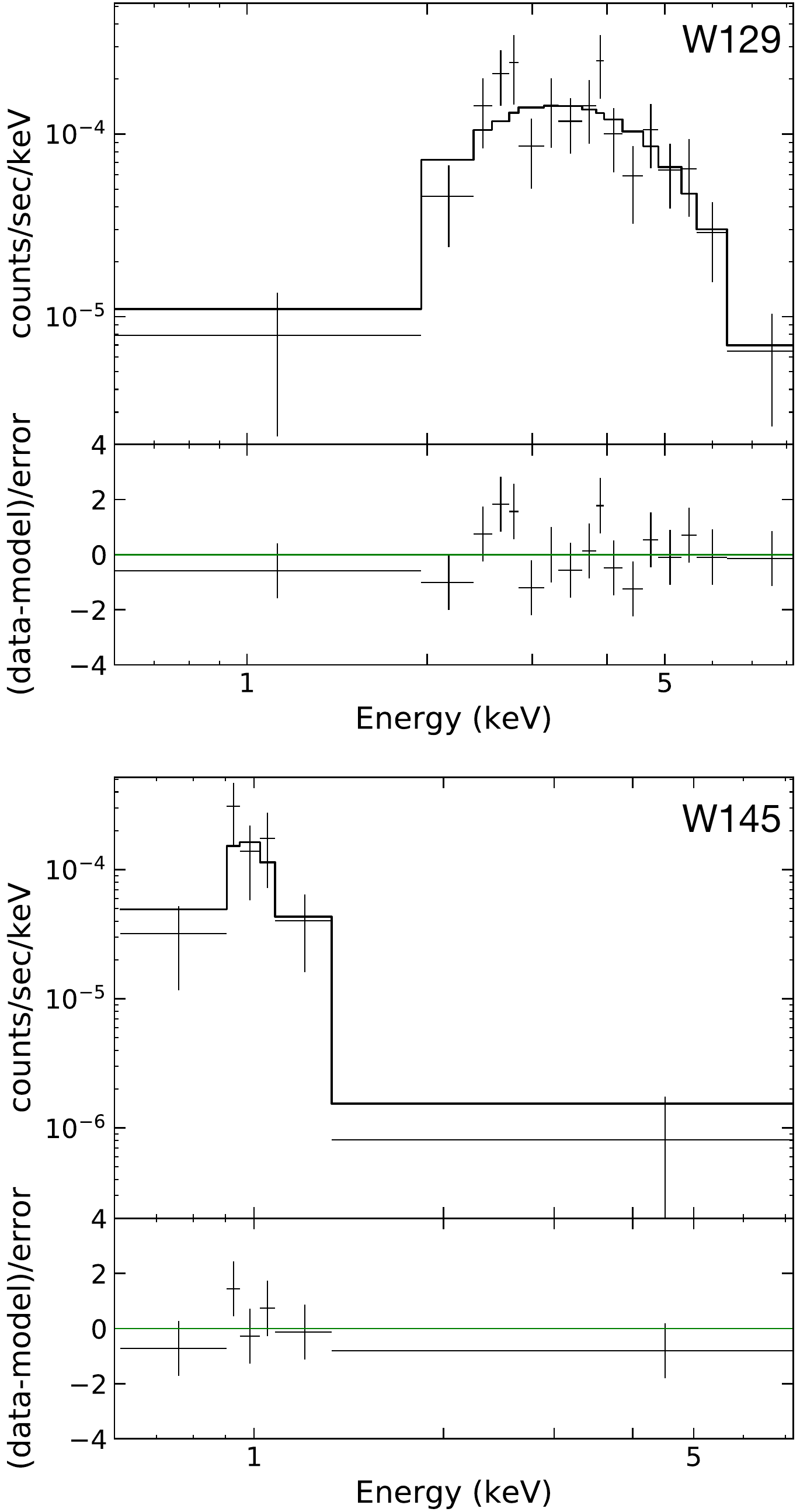}
    \caption{{\it Chandra} ACIS-S spectra of W129 (top) and W145 (bottom) overplotted with best-fitting power-law (W129) and {\tt vmekal} (W145) model. The bottom panel of each plot shows the fitting residuals. The spectra are re-binned only for plotting purpose.}
    \label{fig:spec_w35_w125}
\end{figure}

In principle, most of the very hard ($X_C\lesssim 0$) sources that lack soft counts are very likely background sources (e.g., AGNs) with high values of $N_\mathrm{H}$. This includes W127, W129, W135, W141 and W142. A typical example is W129, which is heavily absorbed and has zero counts in the $0.5$-$1.5~\mathrm{keV}$ band. The best-fitting power-law model (Fig. \ref{fig:spec_w35_w125}) suggests an $N_\mathrm{H}=11.8^{+6.6}_{-4.7}\times 10^{22}~\mathrm{cm^{-2}}$, which is clearly above the cluster value of $N_\mathrm{H}=1.57\times10^{21}~\mathrm{cm^{-2}}$. Moreover, clear detections in the radio bands (Sec. \ref{sec:sec_likely_agns}) further corroborate their AGN nature.

The soft sources ($X_C \gtrsim 1.8$), including W126, W132, W136, W139, W143, W144, W145 and W146, are likely faint ABs or CVs, whose spectra can be fairly reproduced by either a blackbody with $kT_\mathrm{bb}$ between $0.1$-$0.3~\mathrm{keV}$, or a thermal plasma model (e.g., {\tt vmekal} in {\sc xspec}) with typical plasma temperature $\lesssim 2~\mathrm{keV}$. Blackbody fits give $kT_\mathrm{bb}$s and $R_\mathrm{bb}$ similar to those of MSPs, which seems reasonable since MSPs far from the cluster centre have been observed in other core-collapsed GCs (see e.g., \citealt{Colpi02, Forestell14}); however, we found the spectra are more likely reproduced by optically thin low-temperature plasma, as they possess evidence of emission features at low energies (e.g., emission line at $\sim0.8$-$1~\mathrm{keV}$ that overlaps the Fe L-shell emission), which is more common among faint ABs or CVs. As an example, we show the spectrum of W145 fitted with a {\tt vmekal} model in Fig. \ref{fig:spec_w35_w125}, with a clear emission feature at $\sim 1~\mathrm{keV}$.

Finally, sources with somewhat more balanced spectra, including W124, W125, W128, W130, W131, W133, W134, W137, W138, W140, could be either background AGNs or faint CVs. 

We tried to search for optical counterparts with searching radii up to $2P_\mathrm{err}$ for these new sources, using the {\it Gaia} DR2 archive \citep{gaia2016a, gaia2018}. Limited by the low stellar density at the cluster outskirt, detecting multiple objects in the sub-arcsecond error circles might be rare. Indeed, in all searching radii, we found at most $1$ {\it Gaia} source (Tab. \ref{tab:soft_sources}); however, these unique objects are not necessarily real counterparts, as fainter objects within the searching radii might have magnitudes above the {\it Gaia} limit (G-band limiting magnitude $\approx 21$). Nevertheless, non-detections might still indicate a distant nature for the source. In fact, all of the heavily absorbed sources mentioned above have empty searching regions, which further supports their nature as distant objects. 

\begin{figure}
    \centering
    \includegraphics[scale=0.4]{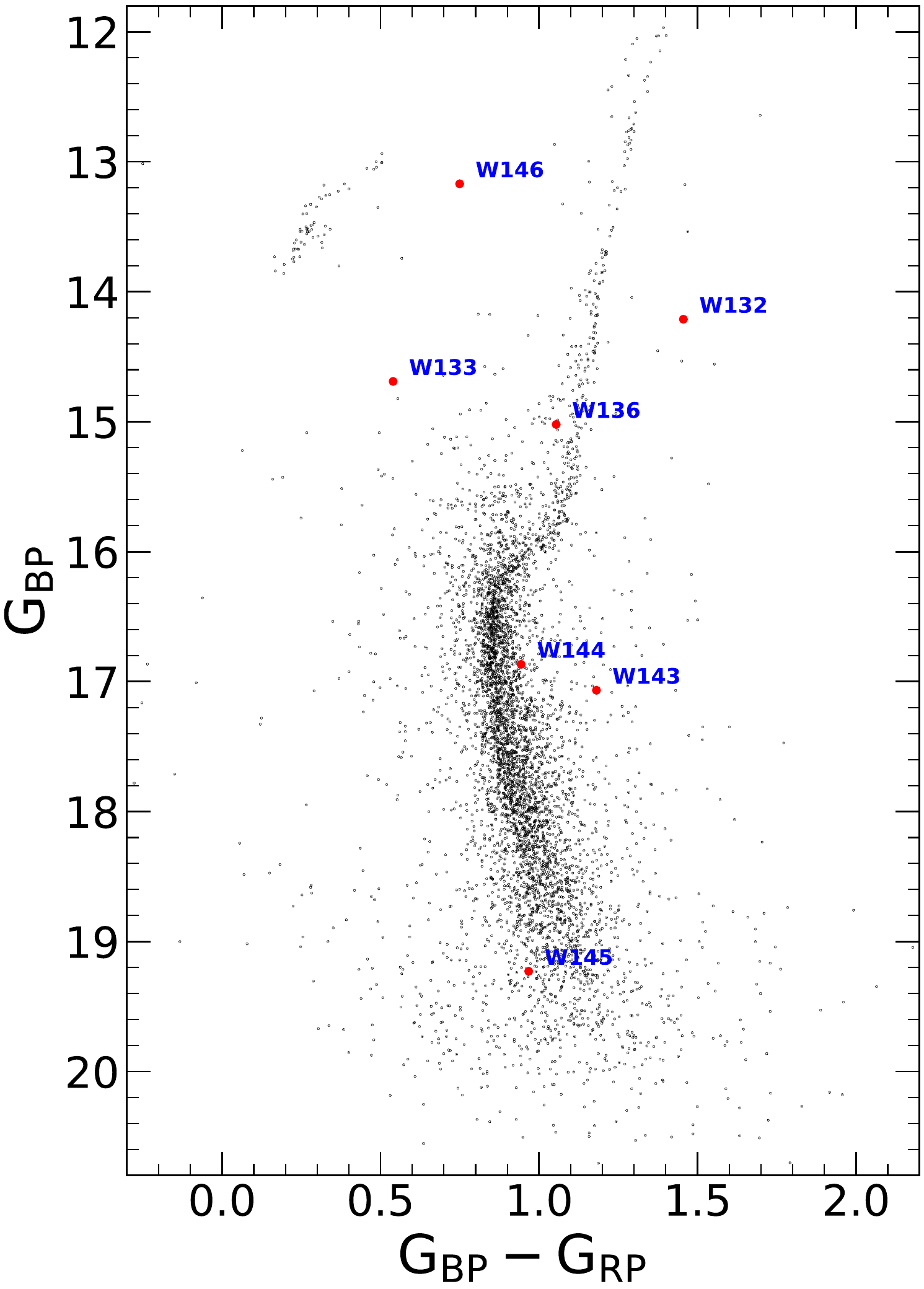}
    \caption{Gaia CMD plotting $\mathrm{G_\mathrm{BP}} - \mathrm{G_\mathrm{RP}}$ colours vs. $\mathrm{G_\mathrm{BP}}$ band magnitudes. The counterparts are indicated with red filled circles.}
    \label{fig:gaia_cmd}
\end{figure}

As complementary information, we check the photometric colours of these potential counterparts using the {\it Gaia} two-band ($\mathrm{G_\mathrm{BP}}$ and $\mathrm{G_\mathrm{RP}}$) magnitudes. We show a {\it Gaia} $\mathrm{G_\mathrm{BP}} - \mathrm{G_\mathrm{RP}}$ CMD in Fig. \ref{fig:gaia_cmd} using stars within a $3\arcmin$ circular region centered on the cluster. We also make use of  proper motion (PM) information from the {\it Gaia} database to confirm membership, only accepting sources that have PMs consistent with the cluster systematic PM ($\mu_\alpha = 3.2908 \pm 0.0026~\mathrm{mas~yr^{-1}},~\mu_\delta = -17.5908\pm 0.0025~\mathrm{mas~yr^{-1}}$; see \citealt{gaia2018b}) at $5\sigma$ as cluster members. 

\begin{figure}
    \centering
    \includegraphics[scale=0.24]{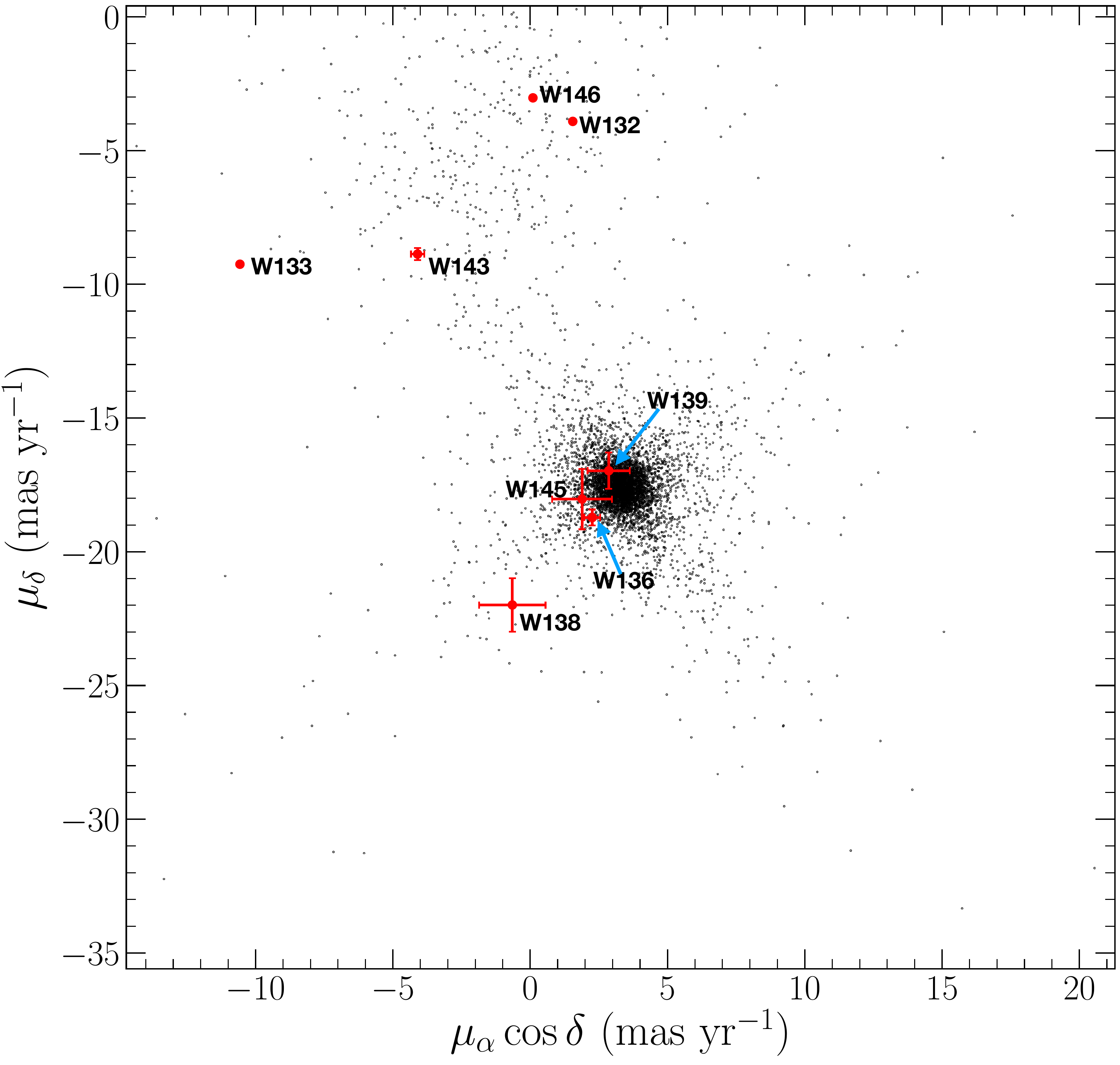}
    \caption{Proper motions (PMs) from {\it Gaia} DR2 of stars within a $3\arcmin$ searching radius. The plot is centered on the cluster proper motion at $\mu_\alpha = 3.2908 \pm 0.0026~\mathrm{mas~yr^{-1}},~\mu_\delta = -17.5908\pm 0.0025~\mathrm{mas~yr^{-1}}$. PM of each counterpart is indicated with red dot and $1~\sigma$ error bars.}
    \label{fig:counterpart_PM}
\end{figure}

There are five sources, W132, W133, W138, W143 and W146, that have clearly inconsistent PMs with the cluster. The red excesses in W132 and W143 (Fig. \ref{fig:gaia_cmd}) are consistent with their foreground nature (as suggested by their distances in Tab. \ref{tab:soft_sources}), while the counterparts to W133 and W146 appear to be bright stars with blue excesses. W132 has a more definite binarity with a {\it Gaia}-measured radial velocity of $|v_r|= 10.11\pm 1.97~\mathrm{km~s^{-1}}$. W133 has a PM clearly discordant with the cluster (Fig. \ref{fig:counterpart_PM}). Intriguingly, W133 is a background source but the counterpart distance is just a few hundred parsecs greater than the cluster (Tab. \ref{tab:soft_sources}). We suspect that W133 is either a halo object or a ``runaway" close binary---likely a CV---ejected by dynamical encounters in the core. The stars found near W138 and W146 could be chance coincidences, as they are relatively far from the X-ray positions.

Three stars in the list are found to have PMs consistent with the cluster value. The PM-selected cluster member W136 has a red giant counterpart, so it could be an RS CVn system. W139 has a definite association with the cluster, but is so faint that no further photometric information is available. W145 has a counterpart consistent with the MS, but is likely to be a chance coincidence, as it lies at $\sim 2P_\mathrm{err}$ off the X-ray position.

\begin{figure}
    \centering
    \includegraphics[scale=0.27]{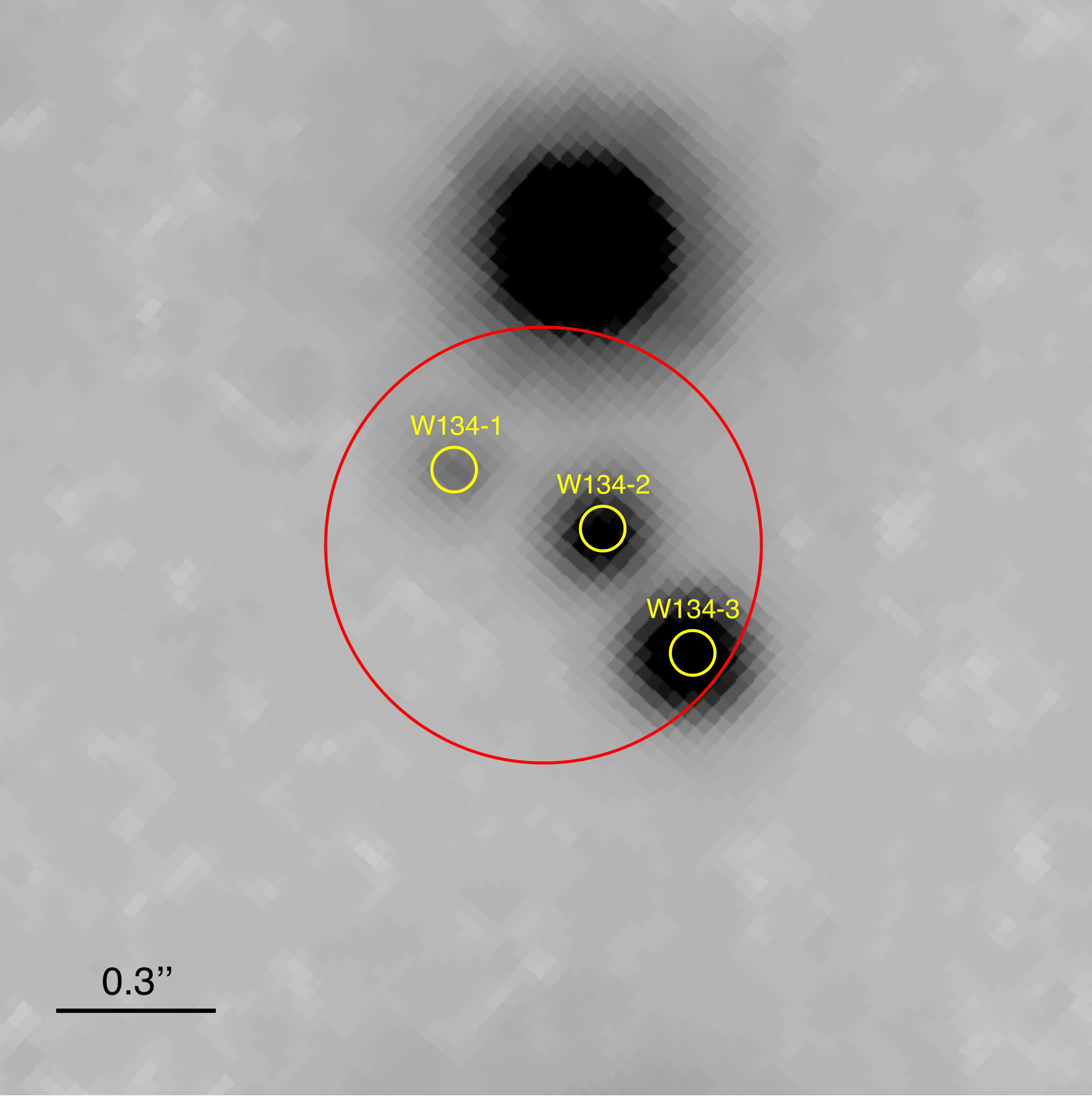}
    \caption{{\it HST/ACS} R625W finding chart for W134 showing a $2\farcs1 \times 2\farcs1$ region centered on the nominal X-ray position of W134, north is up and east is to the left. The red circle indicates the {\it Chandra} 95\% error region while likely counterparts are indicated with yellow circles.}
    \label{fig:finders_w134}
\end{figure}

Finally, W134 is within the {\it HST} field of view, so HUGS photometry is available. We found $3$ stars (Fig. \ref{fig:finders_w134}) with colours consistent with the main sequence (MS) in the $UV_{275}-U_{336}$ and $U_{336} - B_{438}$ CMDs, while in the $V_{606}-I_{814}$ CMD, there is one star (W134-2) that exhibits a very mild red excess relative to the MS (Fig. \ref{fig:hugs_cmds}), which we think is more likely to be the counterpart. W134-2 has a definite cluster membership probability of $98.3\%$ but is less likely a chance coincidence compared to the other two. 





\begin{table}
\centering
    \caption{X-ray catalogue of source within the $2\farcm9$ half-light radius.}
    \resizebox{\columnwidth}{!}{
    \begin{tabular}{ccccccc}
    \toprule
    ID & RA & DEC & $P_\mathrm{err}^a$ & \multicolumn{2}{c}{Counts$^b$} & Flux$^c$ \\
       & \multicolumn{2}{c}{(J2000)} & & $0.5$-$1.5~\mathrm{keV}$ & $1.5$-$6~\mathrm{keV}$ & $0.5$-$6.0~\mathrm{keV}$ \\
    \midrule
    U5   & 17:40:54.539 & $-$53:40:44.912 & 0.33 & 167  & 31   & 41.7   \\
    U7   & 17:40:52.843 & $-$53:41:22.170 & 0.32 & 252  & 150  & 84.6   \\
    U10  & 17:40:48.987 & $-$53:39:48.986 & 0.29 & 283  & 1504 & 390.8  \\
    U11  & 17:40:45.786 & $-$53:40:41.898 & 0.32 & 72   & 55   & 26.7   \\
    U12  & 17:40:44.627 & $-$53:40:41.957 & 0.30 & 382  & 363  & 157.3  \\
    U13  & 17:40:44.093 & $-$53:40:39.506 & 0.31 & 107  & 135  & 51.4   \\
    U14  & 17:40:43.340 & $-$53:41:55.821 & 0.40 & 17   & 10   & 5.7    \\
    U15  & 17:40:42.923 & $-$53:40:34.127 & 0.34 & 42   & 9    & 11.2   \\
    U16  & 17:40:42.659 & $-$53:42:15.698 & 0.42 & 15   & 10   & 5.4    \\
    U17  & 17:40:42.659 & $-$53:40:19.628 & 0.28 & 5358 & 6114 & 2458.0 \\
    U18  & 17:40:42.619 & $-$53:40:27.964 & 0.29 & 995  & 1142 & 470.2  \\
    U19  & 17:40:42.317 & $-$53:40:29.073 & 0.28 & 2777 & 5934 & 1927.3 \\
    U21  & 17:40:41.843 & $-$53:40:21.740 & 0.29 & 1974 & 2279 & 901.4  \\
    U22  & 17:40:41.711 & $-$53:40:29.380 & 0.31 & 104  & 131  & 50.3   \\
    U23  & 17:40:41.605 & $-$53:40:19.624 & 0.28 & 1677 & 5915 & 1609.5 \\
    U24  & 17:40:41.481 & $-$53:40:04.790 & 0.28 & 4797 & 353  & 1088.8 \\
    U25  & 17:40:41.248 & $-$53:40:26.164 & 0.33 & 39   & 49   & 18.7   \\
    U28  & 17:40:38.915 & $-$53:39:51.472 & 0.30 & 351  & 629  & 206.1  \\
    U31  & 17:40:34.213 & $-$53:41:15.588 & 0.37 & 26   & 18   & 9.5    \\
    U41  & 17:40:45.015 & $-$53:39:55.544 & 0.36 & 31   & 4    & 7.5    \\ 
    U42  & 17:40:43.064 & $-$53:38:31.633 & 0.34 & 141  & 15   & 33.0   \\
    U43  & 17:40:40.556 & $-$53:40:23.180 & 0.36 & 25   & 10   & 7.4    \\
    U60  & 17:40:47.825 & $-$53:41:28.649 & 0.41 & 10   & 11   & 4.4    \\
    U61  & 17:40:45.231 & $-$53:40:28.964 & 0.32 & 79   & 85   & 34.8   \\
    U62  & 17:40:30.426 & $-$53:39:17.921 & 0.48 & 12   & 5    & 3.7    \\
    U63  & 17:40:31.680 & $-$53:38:46.731 & 0.42 & 31   & 7    & 8.0    \\
    U65  & 17:40:37.578 & $-$53:39:18.215 & 0.43 & 14   & 3    & 3.6    \\
    U66  & 17:40:38.952 & $-$53:38:49.981 & 0.48 & 9    & 4    & 2.7    \\
    U67  & 17:40:40.074 & $-$53:40:16.880 & 0.41 & 8    & 6    & 3.0    \\
    U68  & 17:40:40.694 & $-$53:38:33.180 & 0.44 & 1    & 20   & 4.4    \\
    U69  & 17:40:40.880 & $-$53:40:17.513 & 0.37 & 15   & 9    & 5.1    \\
    U70  & 17:40:41.705 & $-$53:40:33.608 & 0.32 & 59   & 41   & 22.1   \\
    U73  & 17:40:42.692 & $-$53:39:29.145 & 0.38 & 22   & 8    & 6.3    \\
    U75  & 17:40:43.662 & $-$53:40:30.993 & 0.39 & 9    & 10   & 4.2    \\
    U76  & 17:40:43.827 & $-$53:41:16.679 & 0.38 & 21   & 7    & 5.9    \\
    U77  & 17:40:44.140 & $-$53:42:11.808 & 0.42 & 18   & 7    & 5.2    \\
    U79  & 17:40:46.415 & $-$53:40:04.248 & 0.38 & 18   & 9    & 5.9    \\
    U80  & 17:40:46.445 & $-$53:41:56.863 & 0.39 & 20   & 15   & 7.4    \\
    U81  & 17:40:46.482 & $-$53:41:15.771 & 0.38 & 15   & 11   & 5.5    \\
    U82  & 17:40:48.549 & $-$53:39:39.775 & 0.37 & 31   & 9    & 8.6    \\
    U83  & 17:40:49.636 & $-$53:40:43.350 & 0.46 & 5    & 6    & 2.4    \\
    U84  & 17:40:54.815 & $-$53:40:20.169 & 0.40 & 32   & 5    & 7.8    \\
    U86  & 17:40:37.483 & $-$53:41:47.651 & 0.35 & 56   & 11   & 14.6   \\
    U87  & 17:40:42.888 & $-$53:40:26.772 & 0.37 & 14   & 10   & 5.3    \\
    U88  & 17:40:42.848 & $-$53:40:23.886 & 0.37 & 20   & 8    & 6.2    \\
    U89  & 17:40:43.651 & $-$53:40:24.927 & 0.41 & 11   & 3    & 3.1    \\
    U90  & 17:40:41.808 & $-$53:40:14.924 & 0.36 & 26   & 8    & 7.2    \\
    U91  & 17:40:42.422 & $-$53:40:42.510 & 0.40 & 6    & 9    & 3.3    \\
    U92  & 17:40:43.923 & $-$53:40:35.690 & 0.40 & 11   & 4    & 3.2    \\
    U93  & 17:40:42.361 & $-$53:40:47.315 & 0.42 & 6    & 7    & 2.8    \\
    U94  & 17:40:42.893 & $-$53:40:49.413 & 0.40 & 14   & 2    & 3.4    \\
    U95  & 17:40:40.334 & $-$53:40:44.916 & 0.47 & 4    & 4    & 1.7    \\
    U96  & 17:40:39.090 & $-$53:40:23.414 & 0.42 & 11   & 2    & 2.7    \\
    U97  & 17:40:43.916 & $-$53:40:05.702 & 0.44 & 4    & 6    & 2.2    \\
    U98  & 17:40:41.006 & $-$53:40:58.883 & 0.41 & 12   & 2    & 3.0    \\
    U99  & 17:40:46.439 & $-$53:40:30.767 & 0.50 & 6    & 1    & 1.5    \\
    U100 & 17:40:38.212 & $-$53:40:46.872 & 0.45 & 6    & 4    & 2.1    \\
    U101 & 17:40:45.412 & $-$53:41:01.797 & 0.39 & 19   & 3    & 4.6    \\
    U102 & 17:40:38.877 & $-$53:39:43.446 & 0.40 & 11   & 9    & 4.2    \\
    U103 & 17:40:35.718 & $-$53:40:13.018 & 0.44 & 9    & 3    & 2.5    \\
    U104 & 17:40:43.138 & $-$53:39:29.280 & 0.44 & 5    & 8    & 2.8    \\
    U105 & 17:40:36.535 & $-$53:41:08.319 & 0.44 & 8    & 5    & 2.7    \\
    U106 & 17:40:43.768 & $-$53:39:17.814 & 0.51 & 3    & 5    & 1.7    \\
    U107 & 17:40:34.119 & $-$53:40:17.295 & 0.39 & 16   & 10   & 5.4    \\
    U108 & 17:40:52.105 & $-$53:39:48.404 & 0.46 & 9    & 6    & 3.2    \\
    U109 & 17:40:52.755 & $-$53:40:53.217 & 0.50 & 8    & 3    & 2.3    \\
    U110 & 17:40:33.455 & $-$53:39:17.335 & 0.50 & 8    & 3    & 2.6    \\
    U111 & 17:40:29.884 & $-$53:40:27.207 & 0.45 & 3    & 14   & 3.7    \\
    U112 & 17:40:50.385 & $-$53:39:06.322 & 0.53 & 7    & 3    & 2.1    \\
    U113 & 17:40:42.764 & $-$53:40:21.041 & 0.31 & 161  & 61   & 50.0   \\
    U114 & 17:40:43.480 & $-$53:40:34.662 & 0.45 & 4    & 5    & 2.0    \\
    U116 & 17:40:42.247 & $-$53:40:20.292 & 0.36 & 15   & 14   & 6.3    \\
    U117 & 17:40:42.164 & $-$53:40:25.882 & 0.41 & 7    & 6    & 2.9    \\
    U118 & 17:40:41.587 & $-$53:40:16.202 & 0.43 & 3    & 8    & 2.3    \\
    U119 & 17:40:41.272 & $-$53:40:19.672 & 0.40 & 4    & 11   & 3.2    \\
    U120 & 17:40:46.528 & $-$53:40:15.962 & 0.53 & 6    & 0    & 1.3    \\
    \bottomrule
    \end{tabular}
    }
\label{tab:x_ray_catalog}
\end{table}

\begin{table}
\centering
    \contcaption{X-ray catalogue of sources within the $2\farcm9$ half-light radius.}
    \resizebox{\columnwidth}{!}{
    \begin{tabular}{ccccccc}
    \toprule
    ID  &      RA      &    DEC        & $P_\mathrm{err}$  & \multicolumn{2}{c}{Counts} & Flux \\
        & \multicolumn{2}{c}{(J2000)}  &       &  $0.5$-$1.5~\mathrm{keV}$ & $1.5$-$6~\mathrm{keV}$ & $0.5$-$6.0~\mathrm{keV}$ \\
    \midrule
    U121 & 17:40:33.642 & $-$53:39:35.282 & 0.50 & 2   & 8  & 2.1 \\
    U122 & 17:40:47.914 & $-$53:39:25.152 & 0.58 & 1   & 5  & 1.3 \\
    U123 & 17:40:49.632 & $-$53:38:46.252 & 0.60 & 2   & 6  & 1.7 \\
    W124   & 17:40:30.511 & $-$53:42:34.201 & 0.38 & 103 & 108 & 49.1 \\
    W125   & 17:40:27.976 & $-$53:42:00.059 & 0.39 & 73  & 66 & 30.6  \\
    W126   & 17:40:59.672 & $-$53:40:39.095 & 0.38 & 54  & 10 & 8.1   \\
    W127  & 17:40:58.559 & $-$53:40:14.878 & 0.52 & 3   & 10  & 3.0   \\
    W128  & 17:40:56.736 & $-$53:39:39.863 & 0.33 & 102 & 104 & 45.5  \\
    W129  & 17:40:56.204 & $-$53:39:11.229 & 0.35 & 0   & 123 & 61.6  \\
    W130  & 17:40:50.597 & $-$53:38:25.760 & 0.34 & 103 & 90  & 36.6  \\
    W131  & 17:40:37.690 & $-$53:42:46.340 & 0.42 & 32  & 22  & 9.4   \\
    W132  & 17:40:36.503 & $-$53:42:41.158 & 0.52 & 24  & 1   & 4.5   \\
    W133  & 17:40:58.337 & $-$53:41:53.413 & 0.37 & 78  & 40  & 23.3  \\
    W134  & 17:40:42.036 & $-$53:40:37.877 & 0.41 & 9   & 5   & 2.9   \\
    W135  & 17:40:48.754 & $-$53:43:04.444 & 0.64 & 2   & 9   & 5.4   \\
    W136  & 17:40:53.157 & $-$53:40:58.363 & 0.63 & 5   & 0   & 0.6   \\
    W137  & 17:40:58.548 & $-$53:39:17.142 & 0.42 & 42  & 31  & 14.8 \\
    W138 & 17:40:50.710 & $-$53:43:02.721 & 0.73 & 5   & 5    & 3.0  \\
    W139 & 17:40:29.296 & $-$53:42:21.018 & 0.85 & 17  & 5    & 4.1  \\
    W140 & 17:40:53.868 & $-$53:41:30.826 & 0.57 & 4   & 6    & 1.8  \\
    W141 & 17:40:26.669 & $-$53:41:18.152 & 0.74 & 1   & 16   & 8.8  \\
    W142 & 17:41:00.896 & $-$53:40:52.215 & 0.62 & 2   & 12   & 4.2  \\
    W143 & 17:40:22.994 & $-$53:40:34.913 & 0.72 & 17  & 4    & 3.6  \\
    W144 & 17:40:33.992 & $-$53:39:07.244 & 0.74 & 8   & 2    & 2.3  \\
    W145 & 17:40:27.889 & $-$53:38:45.537 & 0.88 & 22  & 8    & 5.6  \\
    W146 & 17:40:57.386 & $-$53:38:40.507 & 0.65 & 23  & 0    & 2.8  \\
\bottomrule
\multicolumn{7}{l}{$^{a, b}$ $P_\mathrm{err}$ (in $\arcsec$) and counts for U-sources are from \citet{Bogdanov10}. Counts of}\\
\multicolumn{7}{l}{the W sources are calculated by {\tt srcflux}.}\\
\multicolumn{7}{l}{$^c$ Source fluxes are in units of $10^{-16}~\mathrm{erg~s^{-1}~cm^{-2}}$; fluxes for all U sources are from}\\
\multicolumn{7}{l}{\citet{Bogdanov10}, while fluxes for the W sources are model-independent fluxes}\\
\multicolumn{7}{l}{calculated by {\tt srcflux}.}
    \end{tabular}
    }
\end{table}

\begin{table*}
    \centering
    \caption{Optical counterparts to the new sources.}
    \begin{tabular}{ccccccccc}
    \toprule
        ID   & \multicolumn{2}{c}{Optical position} &         Offset$^a$            &   \multicolumn{2}{c}{Magnitudes$^\dag$}     &Membership$^b$ & Dist$^c$ & Comments\\
             &        RA        &         DEC       & ($\arcsec$;~$P_\mathrm{err}$) & $\mathrm{G_{BP}}$ & $\mathrm{G_{RP}}$& & (kpc) \\
    \midrule
        W126  & 17:40:59.662 & $-$53:40:39.001 & $0.13;~0.34$ &    -    &    -    &      -     & - & Faint AB or CV? \\
        W131  & 17:40:37.686 & $-$53:42:46.324 & $0.04;~0.10$ &    -    &    -    &      -     & - & Faint AB or CV?\\
        W132  & 17:40:36.481 & $-$53:42:41.151 & $0.20;~0.38$ & $14.21$ & $12.76$ & Foreground & $0.257^{+0.002}_{-0.002}$ & Foreground AB?\\
        W133  & 17:40:58.332 & $-$53:41:53.318 & $0.11;~0.29$ & $14.69$ & $14.15$ & Background & $2.84^{+0.31}_{-0.26}$ & Runaway CV? \\
        W134  & 17:40:42.024 & $-$53:40:37.854 & $0.11;~0.27$ & $20.30$ & $19.10$ &   Member   & $1.82^{+0.38}_{-0.27}$ & Faint AB or CV? \\
        W136  & 17:40:53.174 & $-$53:40:58.642 & $0.32;~0.51$ & $15.02$ & $13.97$ &   Member   & $1.57^{+0.72}_{-0.38}$ & RS CVn?\\
        W138 & 17:40:50.856 & $-$53:43:03.335 & $1.43;~1.96$ &    -    &    -    & Foreground & $0.58^{+0.76}_{-0.21}$ & - \\
        W139 & 17:40:29.293 & $-$53:42:21.196 & $0.18;~0.21$ &    -    &    -    &   Member   & $3.59^{+3.32}_{-1.79}$ & Faint AB or CV?\\
        W143 & 17:40:22.925 & $-$53:40:34.066 & $1.04;~1.45$ & $17.07$ & $15.89$ & Foreground & $1.36^{+0.40}_{-0.30}$ & Foreground AB?\\
        W144 & 17:40:33.890 & $-$53:39:08.198 & $1.32;~1.78$ & $16.86$ & $15.92$ &      -     & - & - \\
        W145 & 17:40:27.722 & $-$53:38:45.047 & $1.56;~1.78$ & $19.23$ & $18.26$ &   Member   & $3.13^{+3.30}_{-1.83}$ & Faint AB or CV? \\
        W146 & 17:40:57.314 & $-$53:38:39.885 & $0.89;~1.37$ & $13.17$ & $12.42$ & Foreground & $0.70^{+0.02}_{-0.02}$ & Faint AB or CV?\\
    \bottomrule
    \multicolumn{8}{l}{$^a$Offsets from the X-ray positions in terms of $\arcsec$ and $P_\mathrm{err}$}\\
    \multicolumn{8}{l}{$^b$Cluster membership determinations based on {\it Gaia} proper motions and/or distances or, for W134,} \\
    \multicolumn{8}{l}{the membership probability from the HUGS data}\\
    \multicolumn{8}{l}{$^c$Distance estimates according to \citet{Bailer-Jones18}. The reported errors are at the}\\
    \multicolumn{8}{l}{$68\%$ confidence level}\\
    \multicolumn{8}{l}{$^\dag$Magnitudes in the {\it Gaia} $G_\mathrm{BP}$ and $G_\mathrm{RP}$ band passes; for W134, the first and second column}\\
    \multicolumn{8}{l}{corresponds to magnitudes at the {\it HST}/ACS $F606W$ and $F814W$ filter, respectively}
    \end{tabular}
    \label{tab:soft_sources}
\end{table*}

\bsp	
\label{lastpage}
\end{document}